\documentclass[11pt]{article}

\usepackage[final]{acl}

\usepackage{times}
\usepackage{latexsym}

\usepackage[T1]{fontenc}

\usepackage[utf8]{inputenc}

\usepackage{microtype}

\usepackage{inconsolata}

\usepackage{graphicx}

\usepackage{microtype}
\usepackage{hyperref}
\usepackage{url}
\usepackage{booktabs}
\usepackage{multirow}
\usepackage{amssymb} %
\usepackage{pifont}  %
\usepackage{enumitem}
\usepackage{wrapfig}
\usepackage{subcaption}
\usepackage{amsmath}

\usepackage{lineno}

\usepackage{xcolor}

\newcommand{\seiji}[1]{#1}
\newcommand{\cameraready}[1]{#1}

\definecolor{darkblue}{rgb}{0, 0, 0.5}
\hypersetup{colorlinks=true, citecolor=darkblue, linkcolor=darkblue, urlcolor=darkblue}

\title{Align Then Adapt: Label-Efficient Adapter Learning for \\ Asymmetric Dense Retrieval}

\author{
    Seiji Maekawa \\
    Megagon Labs \\
    \texttt{seiji@megagon.ai} \\
    \footnotesize{0000-0003-4283-0929}\\\And
    Moin Aminnaseri \\
    Megagon labs \\
    \texttt{moin@megagon.ai} \\
    \footnotesize{0009-0003-1665-9678}\\\And
    Pouya Pezeshkpour \\
    Megagon labs \\
    \texttt{pouya@megagon.ai} \\
    \footnotesize{0009-0002-7055-0035}\\\And
    Estevam Hruschka \\
    Megagon labs \\
    \texttt{estevam@megagon.ai} \\
    \footnotesize{0000-0003-1499-2808}\\
    }

\definecolor{deepgreen}{rgb}{0.0, 0.5, 0.0}

\newcommand{\xmark}{\textcolor{red}{\textbf{\ding{55}}}}
\newcommand{\cmark}{\textcolor{deepgreen}{\textbf{\ding{51}}}}

\begin{document}
\maketitle
\begin{abstract}
Dense retrieval systems increasingly face an asymmetry between complex instruction-like queries and relatively simple, static document collections. While stronger embedders can better understand such queries, re-embedding large corpora or fine-tuning large models is often impractical. We propose Efficient Retrieval Adapter (ERA), a query-side adapter learning framework for re-index-free retrieval adaptation. ERA first aligns the embedding spaces of a strong query embedder and a lightweight document embedder using unlabeled corpus documents, and then adapts the aligned query representation with a small number of labeled query-document pairs. Across 126 MAIR retrieval tasks from six domains, ERA improves average nDCG@10 by up to 8.2 points in symmetric settings and by more than 12 points in asymmetric settings, while using substantially fewer labels than supervised adapter training. These results show that retrieval systems can benefit from stronger query understanding without updating backbone embedders or rebuilding document indexes.

\end{abstract}

\section{Introduction}
\label{sec:intro}
\begin{figure}
   \centering
   \includegraphics[width=\linewidth]{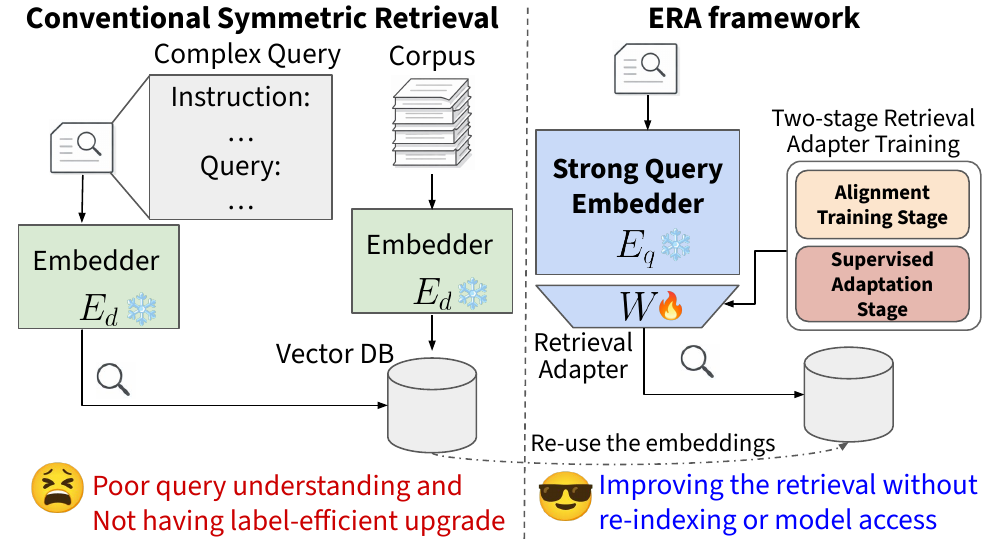}
   \caption{ERA enables re-index-free asymmetric dense retrieval by aligning a strong query embedder with a lightweight document index. Unlike conventional symmetric retrieval, ERA trains a lightweight query-side adapter in two stages: self-supervised alignment using unlabeled corpus documents and supervised adaptation using limited labeled query-document pairs.}
   \label{fg:motivation}
\end{figure}

Dense retrieval systems \citep{karpukhin-etal-2020-dense} are increasingly expected to handle queries that go far beyond short keyword-like inputs. 
In many realistic scenarios, users express complex intent through long instructions, multi-step constraints, or task-specific descriptions, while the target documents remain comparatively simple and static. 
Recent studies \citep{sun-etal-2024-mair, weller-etal-2025-followir,elsen2025nevir} have introduced benchmarks that reflect this asymmetry, where queries are often long and instruction-like.
This asymmetry creates a fundamental mismatch in retrieval: understanding the query may require strong reasoning and instruction-following ability, whereas representing the document collection efficiently demands a lightweight indexing pipeline.
However, most existing retrieval systems treat the query and document sides symmetrically, using the same model, the same adaptation strategy, or the same optimization assumptions for both, as illustrated in the left part of Figure \ref{fg:motivation}. 
As a result, they either fail to account for the complexity and semantic gaps between queries and documents, or do so only at the substantial cost of fine-tuning.

A natural solution would be to fine-tune a LLM-based retrieval model \citep{zhang2025qwen3embedding,lee2025nvembed} so that it can better follow complex instructions. 
In practice, however, this is difficult to deploy. 
Fine-tuning large embedding models is computationally expensive, memory-intensive, and operationally burdensome, especially when multiple tasks or domains must be supported. 
Beyond training cost, supervised adaptation itself depends on collecting query-document relevance labels, which is often the real bottleneck. Such labels are expensive to create, highly task-dependent, and difficult to scale across domains. 
Recent work \citep{chromadb2024embedding,yoon-etal-2024-search} has explored lightweight alternatives to fine-tuning, such as training retrieval adapters. 
However, these methods still require a large number of labeled examples per task and do not effectively address the representation gap between different embedding models or the semantic gap between complex queries and simple documents. 
Moreover, empirical evidence \citep{chromadb2024embedding} suggests that adapter-based training alone may not yield noticeable gains unless each task has at least around 1,500 labeled query-document pairs, making low-resource adaptation challenging even when full fine-tuning is avoided.

In many realistic settings, however, practitioners do not have enough labeled data to support heavy supervised adaptation, nor the computational budget to repeatedly fine-tune large embedders or re-index a potentially large-volume document corpus. 
This motivates a different question: \textit{can retrieval be efficiently adapted to complex queries without updating the backbone model and without re-indexing the potentially large-volume document corpus?}
To the best of our knowledge, addressing query-document complexity mismatch using a lightweight retrieval adapter alone, while keeping the underlying embedding models frozen, has not been systematically studied.

In this work, we propose Efficient Retrieval Adapter (ERA), a novel framework for training retrieval adapters in a label-efficient manner.
Inspired by the pre-training and supervised fine-tuning stages of LLMs, the key idea of ERA is to decouple retrieval adapter training into two stages: a self-supervised alignment stage and a supervised adaptation stage.
In the alignment stage, we train the retrieval adapter to align the embedding spaces between a query embedder and a document embedder using self-supervised learning on unlabeled data.
This step reduces the semantic gap between the two embedding spaces, making subsequent supervised training more effective and efficient, allowing the retrieval adapter to achieve performance gains even with fewer labeled examples, as observed in LLM fine-tuning after pre-training \citep{devlin-etal-2019-bert}.
In the supervised adaptation stage, we tune the retrieval adapter using a small number of labeled query-document pairs, capturing semantic nuances between the query and document embedding spaces that are not fully captured by the alignment stage.
This design avoids the prohibitive cost of re-embedding the corpus with a large model, while allowing the system to benefit from richer query understanding at inference time.

The key contributions are as follows:
\begin{itemize}[leftmargin=.75em]
   \item We propose ERA, a novel framework for training retrieval adapters in a label-efficient manner, which decouples the training into a self-supervised alignment stage and a supervised adaptation stage.
   \item We conduct extensive experiments with five embedding models from three model families on the MAIR benchmark \citep{sun-etal-2024-mair}, spanning 126 retrieval tasks across six domains.  
   \seiji{We show that ERA, despite being a linear adapter, effectively combines stronger query embedders with weaker document embedders, including across different model families, by bridging both the representation gap between embedding models and the semantic gap between complex queries and simple documents, improving average \textbf{nDCG@10 by up to 12\%} over zero-shot retrieval of the document embedder without expensive corpus re-indexing. 
   \item We observe that the alignment stage significantly improves the label efficiency of adapter training, allowing ERA to achieve substantial performance gains with fewer than 100 labeled examples per task, which is a significant reduction compared to prior methods that require around 1,500 labeled examples for noticeable gains.}
   \item We analyze the robustness of ERA against out-of-domain tasks and its hyperparameters, showing that it is a practical solution for real-world retrieval scenarios. 
\end{itemize}

\begin{table*}[t]
\begin{center}
   \scalebox{0.7}{
      \begin{tabular}{lcccccc}
      \toprule
                    & \textbf{No Parameter} & {\bf $<1000$} &\multicolumn{1}{c}{\bf Asymmetric} & {\bf Domain} & {\bf Adaptation}\\
      {\bf Methods} & \textbf{Access} & {\bf Label Train} &\multicolumn{1}{c}{\bf Embeddings} & {\bf Generality} & {\bf Signal}\\
      \midrule
      Zero-shot Retrieval                             & \cmark & \xmark & \xmark & \cmark & \xmark \\
      Embedding Adapter \citep{chromadb2024embedding} & \cmark & \xmark & \xmark & \xmark & Relevance labels\\
      Search Adaptor \citep{yoon-etal-2024-search}    & \cmark & \xmark & \xmark & \xmark & Relevance labels\\
      Drift-Adapter \citep{vejendla-2025-drift}       & \cmark & \xmark & \cmark & \cmark & Alignment Only\\
      \midrule
      \textbf{ERA} (Ours)                                      & \cmark & \cmark & \cmark & \cmark & Alignment + Relevance\\
      \bottomrule
      \end{tabular}
   }
\end{center}
\caption{Comparison of the proposal, ERA, with existing methods. \seiji{``Adaptation Signal'' summarizes the supervision used to train the adapter. ERA combines cross-embedder alignment with relevance-aware supervised adaptation.} }
\label{tab:comparison}
\end{table*}

\section{Related Work}
\label{sec:related}
\noindent \textbf{LLM Embedders. }
While traditional BERT-based embedders \citep{karpukhin-etal-2020-dense,chen-etal-2024-m3} have long been the standard for retrieval, recent LLM-based embedders \citep{lee2025nvembed,zhang2025qwen3embedding} often achieve stronger performance due to their superior reasoning and instruction-following abilities.
However, these gains come at a substantial cost: LLM-based embedders are expensive to fine-tune and deploy, making them less practical for real-world settings that must support many tasks or domains, especially given their large model sizes (e.g., 4B+ parameters).

\noindent \textbf{Embedding Adapters.}
While prior work has explored fine-tuning for retrieval improvement \citep{kim-baek-2025-syntriever,dai2023promptagator,schlatt2025distillm}, Embedding Adapters \citep{chromadb2024embedding} and Search Adaptor \citep{yoon-etal-2024-search} provide lightweight alternatives to full fine-tuning, but remain limited in our setting: both rely on substantial task-specific supervision and require at least 1,500 labeled examples per task for noticeable gains.
Moreover, neither is designed to bridge the representation gap between heterogeneous query and document embedders. Drift-Adapter \citep{vejendla-2025-drift} shows that adapters can align embedding spaces across versions of the same model for re-index-free retrieval adaptation, but does not study supervised label-efficient retrieval improvement. Overall, although embedding adapters are promising for retrieval adaptation, as summarized in Table \ref{tab:comparison}, their label efficiency and ability to bridge both cross-model representation gaps and query-document semantic gaps remain underexplored.

\noindent \textbf{Transferability of Embeddings.}
Recent work suggests that representations from different models share meaningful geometric structure. \citet{chen2025transferring} show that affine stitches can transfer linear features across language models, \citet{huh2024position} argue that representations increasingly converge as models scale, and \citet{jha2025harnessing} show that embeddings from different text encoders can be translated through a shared latent structure even without paired data. However, these studies mainly support the assumption that cross-model geometry is transferable, rather than showing how to use it for retrieval adaptation under query-document asymmetry.

\begin{figure*}[t]
\begin{center}
\includegraphics[width=0.8\linewidth]{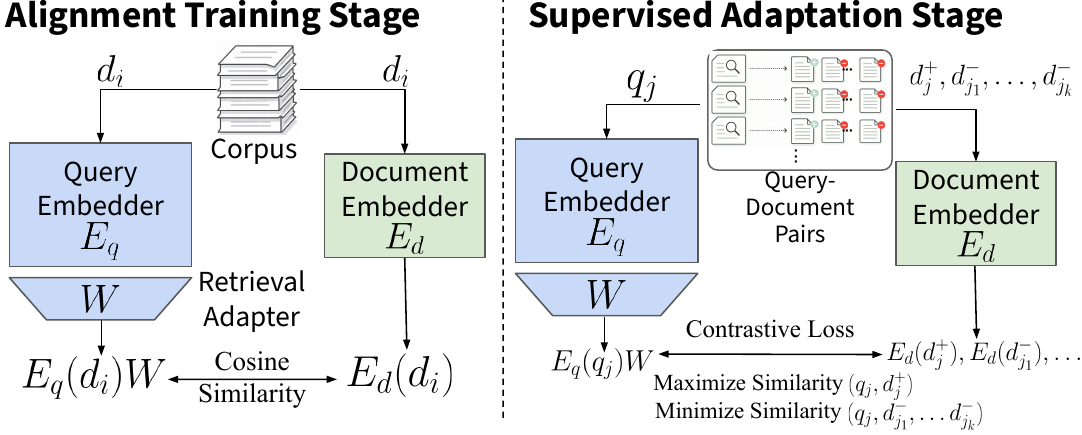}
\end{center}
\caption{The overview of Efficient Retrieval Adapter. In the alignment stage, the same unlabeled document is encoded by both embedders and used as a positive pair. In the adaptation stage, labeled query-document pairs and hard negatives train the same adapter for retrieval.}
\label{fg:ERA}
\end{figure*}

\section{Efficient Retrieval Adapter, ERA}
\label{sec:method}
To address the challenges of retrieval adaptation under query-document complexity asymmetry, we propose Efficient Retrieval Adapter (ERA), a novel framework for training retrieval adapters in a label-efficient manner.
In this section, we first formulate the problem of retrieval with an adapter, and then describe the design principles of ERA, followed by our training recipe. 

\subsection{Task Formulation}
\label{ssec:task_formulation}
\noindent \textbf{Zero-shot Retrieval.}
In a traditional retrieval setting \citep{karpukhin-etal-2020-dense,maekawa-etal-2024-retrieval,taguchi-etal-2025-efficient}, we have an embedding model $E$ that maps a query $q$ and a document $d$ to their respective embeddings, $E(q) \in \mathbb{R}^h$ and $E(d) \in \mathbb{R}^h$, where $h$ is the embedding dimension. 
The retrieval process typically involves computing the similarity between the query and document embeddings $\text{sim}(E(q), E(d))$, often using cosine similarity, to rank documents based on their relevance to the query.

\noindent \textbf{Retrieval with Adapter.}
Embedding Adapter \citep{chromadb2024embedding} introduces a lightweight adapter module $\mathbf{W} \in \mathbb{R}^{h \times h}$ that transforms the query embedding space. The adapted query embedding is computed as $E(q)\mathbf{W}$, which is then L2-normalized, and the retrieval process is based on the similarity $\text{sim}(E(q)\mathbf{W}, E(d))$. This allows for task-specific adaptation without updating the full embedding model, but it typically requires a large number of labeled examples to train effectively.

\noindent \textbf{Asymmetric Retrieval with Adapter.}
In our setting, we consider a more general scenario where the query and document embeddings are generated by different models, denoted as $E_q$ and $E_d$, respectively. The retrieval process is based on the similarity $\text{sim}(E_q(q)\mathbf{W}, E_d(d))$, where $\mathbf{W} \in \mathbb{R}^{h_q \times h_d}$ is the retrieval adapter that transforms the query embedding space to better align with the document embedding space.
This formulation allows for asymmetric embedding spaces, which is more realistic in scenarios where queries require powerful reasoning models while documents are indexed with lightweight encoders.

\subsection{Design Principles}
\label{ssec:design_principles}
ERA is designed with the following principles:
\begin{itemize}[leftmargin=.75em]
   \item \textbf{No Parameter Access}: ERA requires no access to the base embedding model parameters, making it compatible with black-box APIs. 
   \item \textbf{Low Label Requirement}: ERA is designed to be effective with fewer than 100 labeled query-document pairs per task, making it practical for real-world scenarios where labeled data is scarce. 
   \item \textbf{Asymmetric Embeddings}: ERA supports heterogeneous encoder pairs, including embedders from entirely different model families, reflecting realistic deployments where queries demand powerful reasoning models while documents are efficiently indexed with lightweight encoders.
   \item \textbf{Domain Generality}: ERA is designed to be effective across a wide range of domains and tasks, without requiring domain-specific adaptations or assumptions.
\end{itemize}

To achieve these design goals, ERA uses a retrieval adapter, a lightweight module, to transform the query embedding space toward the document embedding space. ERA trains this adapter in two stages: a \textit{self-supervised alignment stage} with no supervised labels and a \textit{supervised adaptation stage} with limited labeled data, as illustrated in Figure \ref{fg:ERA}. 
In the alignment stage, we train the retrieval adapter on unlabeled data to align the embedding spaces of the query and document embedders. This helps bridge the representation gap between the two asymmetric spaces, reduces the semantic gap, and makes subsequent supervised training more effective and efficient. In the adaptation stage, we fine-tune the adapter on limited labeled query-document pairs to capture semantic nuances not fully addressed during alignment, improving understanding of complex queries and retrieval performance without re-indexing the corpus.

\subsection{Training Recipe}
\textbf{Alignment Training Stage.}
\label{ssec:alignment_stage}
The core idea of the alignment training stage is to match the embeddings from the query embedder and the document embedder by using identical input documents. 
Specifically, we first collect a large set of unlabeled documents from the target corpus. For each document $d$, we compute its embedding using the query embedder, $E_q(d)$ and its embedding using the document embedder, $E_d(d)$. 
Then, we train the retrieval adapter $\mathbf{W}$ to maximize the similarity between the adapted query-encoder representation of the document and the document embedding, i.e., $\text{sim}(E_q(d)\mathbf{W}, E_d(d))$. We use cosine similarity as the similarity function $\text{sim}$ in our experiments since it is widely-used and employed in our retrieval process.
This training process encourages the retrieval adapter to learn a transformation that aligns the query embedding space with the document embedding space, effectively bridging the representation gap between the two models.

\noindent \textbf{Supervised Adaptation Stage.}
\label{ssec:adaptation_stage}
After the alignment training stage, we fine-tune the retrieval adapter using a small number of labeled query-document pairs.
Specifically, we have a set of labeled examples $\{(q_i, d_i^+, \{d_{i,j}^-\}_{j=1}^k)\}_{i=1}^N$, where $q_i$ is a query, $d_i^+$ is a relevant document, and $\{d_{i,j}^-\}_{j=1}^k$ are irrelevant documents. We compute the adapted query embedding as $E_q(q_i)\mathbf{W}$, the positive document embedding as $E_d(d_i^+)$, and the negative document embeddings as $\{E_d(d_{i,j}^-)\}_{j=1}^k$. 
Then, we optimize the adapter $\mathbf{W}$ to maximize the similarity between the adapted query embedding and the positive document embedding while minimizing the similarity with the negative document embeddings. 
This can be achieved using a contrastive loss function, such as the InfoNCE loss \citep{oord2018representation} or the triplet loss \citep{reimers-gurevych-2019-sentence}. 
This supervised training process allows the adapter to capture the semantic nuances between the query and document embedding spaces that are not captured by the alignment stage.

\noindent \textbf{Domain Generality.}
To support general-domain coverage, we train a single retrieval adapter across all tasks and domains. Specifically, we sample training examples from the full dataset in both the alignment and adaptation stages, enabling the adapter to capture shared structure across diverse retrieval scenarios without requiring domain-specific adaptation.

\begin{table*}[t]
  \centering
  \small
  \begin{tabular}{llrrrrrrr}
    \toprule
    Base model & Variant & \textbf{Avg} & Academic & Code & Finance & Legal & Medical & Web \\
    \midrule
    \multirow{6}{*}{BGE-M3} & Zero-shot & 34.03 & 34.81 & 34.85 & 52.44 & 37.01 & 25.25 & 33.28 \\
     & Embedding Adapter & 26.18 & 21.66 & 30.44 & 42.09 & 25.24 & 18.76 & 26.55 \\
     & Sym.\ ERA w/o adapt. & 34.17 & 35.01 & 33.79 & 53.58 & 38.41 & 25.54 & 33.35 \\
     & Sym.\ ERA & 42.14 & 40.28 & 40.25 & \textbf{60.77} & 41.78 & 32.35 & 44.08 \\
     & Asym.\ ERA w/o adapt. & 43.69 & 39.15 & 40.16 & 58.66 & 40.65 & 34.49 & 47.84 \\
     & \textbf{Asym.\ ERA} & \textbf{46.59} & \textbf{42.50} & \textbf{46.19} & 59.27 & \textbf{45.20} & \textbf{38.24} & \textbf{49.28} \\
    \midrule
    \multirow{6}{*}{OpenAI-small} & Zero-shot & 36.74 & 35.62 & 41.35 & 50.36 & 31.77 & 30.64 & 36.67 \\
     & Embedding Adapter & 29.67 & 23.68 & 34.10 & 38.17 & 25.34 & 26.64 & 30.66 \\
     & Sym.\ ERA w/o adapt. & 37.19 & 36.32 & 41.01 & 50.13 & 30.79 & 31.09 & 37.71 \\
     & Sym.\ ERA & 44.95 & 42.34 & 48.33 & \textbf{56.88} & 37.16 & 38.02 & 46.87 \\
     & Asym.\ ERA w/o adapt. & 46.63 & 41.69 & 49.48 & 52.68 & 34.72 & 40.40 & 50.86 \\
     & \textbf{Asym.\ ERA} & \textbf{49.10} & \textbf{45.45} & \textbf{52.81} & 54.73 & \textbf{39.75} & \textbf{42.34} & \textbf{52.40} \\
    \midrule
    \multirow{6}{*}{Qwen3-0.6B} & Zero-shot & 50.37 & 48.48 & 58.34 & 60.56 & 48.58 & 44.55 & 49.17 \\
     & Embedding Adapter & 38.94 & 34.49 & 48.68 & 51.02 & 35.15 & 32.51 & 38.26 \\
     & Sym.\ ERA w/o adapt. & 51.12 & 49.28 & 57.28 & 62.17 & 49.85 & 45.53 & 50.20 \\
     & Sym.\ ERA & 52.19 & 49.41 & 58.49 & 63.57 & \textbf{49.99} & \textbf{46.28} & 51.76 \\
     & Asym.\ ERA w/o adapt. & 51.44 & 50.11 & 57.53 & 63.13 & 48.35 & 44.70 & 51.07 \\
     & \textbf{Asym.\ ERA} & \textbf{53.33} & \textbf{50.72} & \textbf{59.14} & \textbf{63.86} & 49.66 & 45.68 & \textbf{54.06} \\
    \midrule
    \multirow{4}{*}{Qwen3-8B} & Zero-shot & 57.91 & 56.94 & 61.85 & 69.38 & 57.89 & \textbf{53.75} & 56.65 \\
     & Embedding Adapter & 46.30 & 43.23 & 53.44 & 58.26 & 42.75 & 37.12 & 47.01 \\
     & Sym.\ ERA w/o adapt. & 58.51 & 58.85 & 61.53 & 70.03 & 58.25 & 53.39 & 57.56 \\
     & Sym.\ ERA & \textbf{59.38} & \textbf{59.13} & \textbf{63.03} & \textbf{71.83} & \textbf{58.58} & 53.50 & \textbf{58.63} \\
    \bottomrule
  \end{tabular}
  \caption{nDCG@10 (\%) on MAIR using 20\% labeled query-document pairs. \textbf{Bold} indicates the best result per model and domain. Embedding Adapter and Symmetric ERA use the same model for both query and document embedders. Asymmetric ERA uses Qwen3-8B as the query embedder and the base model as the document embedder.}

  \label{tab:main_results}
\end{table*}

\section{Experiments}
\label{sec:exp}
We apply ERA to conduct extensive experiments to explore the following research questions: 
\noindent 
\textbf{(RQ1)} \textit{Can we effectively combine a stronger query embedder with a weaker document embedder using ERA, and can query and document embedders come from different model families?} 
\textbf{(RQ2)} \textit{How does ERA perform under different label resource settings?} 
\textbf{(RQ3)} \textit{Is a general adapter obtained by ERA effective across different domains?} 
\textbf{(RQ4)} \textit{How sensitive is ERA to key design and training choices?}

\subsection{Setup}
\label{ssec:setup}

\noindent \textbf{Embedders.}
We use the Qwen3-embedding-8B/0.6B \citep{zhang2025qwen3embedding}, OpenAI embeddings \citep{openai2024embedding}, and BGE-M3 \citep{chen-etal-2024-m3} as the base embedders for all methods, including ERA and baselines. We refer to these models as Qwen3-8B/-0.6B, OpenAI-small, and BGE-M3. The details are summarized in Appendix \ref{app:embedders}.

\noindent \textbf{Datasets.}
We evaluate ERA and the baselines on the MAIR benchmark \citep{sun-etal-2024-mair}, which contains 126 retrieval tasks across six domains: Academic, Code, Finance, Legal, Medical, and Web. Built on widely used benchmarks such as BEIR \citep{thakur2021beir} and KILT \citep{petroni-etal-2021-kilt}, MAIR provides broad coverage across domains and tasks. At the same time, it contains only about 80 labeled examples per task on average (10,038 queries in total), making it an ideal testbed for evaluating label efficiency. 
To simulate explicit user intent, we use the queries with instructions in the MAIR benchmark. 
We split labeled examples into 5\%–40\% for training (roughly 500–4,000 queries across all tasks), 10\% for validation, and 50\% for testing. As we vary the training label percentage, we keep the validation and test sets fixed for fair comparison across label settings.

\noindent \textbf{Methods.}
\seiji{
We compare methods under the constraints targeted by ERA: frozen backbone embedders, a fixed document index, and limited labeled query-document pairs. These constraints exclude full fine-tuning, LoRA-style adaptation, and document-side adaptation such as Search Adaptor from our primary comparison.
We instantiate them as follows. \textbf{Zero-shot Retrieval} uses the frozen document embedder without adaptation. \textbf{Embedding Adapter} trains a query-side linear adapter using only labeled pairs, serving as the closest supervised-only fixed-index baseline\footnote{Embedding Adapter corresponds to ERA without the alignment stage.}. \textbf{ERA w/o adapt.} uses only self-supervised alignment on unlabeled corpus documents. \textbf{ERA} is the full method, combining self-supervised alignment with supervised adaptation.
In addition, to evaluate the effectiveness of asymmetric embedding spaces, we compare \textbf{Symmetric ERA}, where the same model is used for both query and document encoding, with \textbf{Asymmetric ERA}, where a stronger model (Qwen3-8B) is used for query encoding and a weaker model is used for document encoding.
}

\noindent \textbf{Training Details.}
For the alignment stage, we sample 1,000 unlabeled documents from the corpus of each training task\seiji{, which corresponds to only $5.2\%$ of the full MAIR corpus in terms of token cost. We report the detailed token cost analysis in Appendix \ref{app:ssec:token_cost_analysis}}.   
We train the retrieval adapter for 100 epochs with a learning rate of 1e-3, a weight decay of 1e-2, and a batch size of $256$.
We then train the adapter on labeled query-document pairs in the adaptation stage for up to 1,000 epochs with early stopping (patience of 5 epochs based on validation loss), a learning rate of 1e-5, a weight decay of 1e-4, and a batch size of $256$. 
For negative sampling, we use TopK-PercPos \citep{gabriel2025hardnegative}, which selects hard negatives based on retrieval scores after the alignment stage. 
We also evaluate naive top-k negative sampling and random negative sampling as ablations.\footnote{For Embedding Adapter, we use random negatives because, without alignment, hard negatives are less effective for bridging the representation gap.} 
We provide more details about training setup in Appendix \ref{sssec:training_hyperparameters}. 

\subsection{Main Results (RQ1)}
\label{ssec:results}

\seiji{
Table \ref{tab:main_results} shows the nDCG@10 results of Symmetric/Asymmetric ERA and baselines on the MAIR benchmark using 20\% of labeled query-document pairs for training.
We report a widely used retrieval metric, nDCG@10, as the main metric in the paper. 
We also evaluate other training ratios in Appendix \ref{sssec:main_results_varying_train_ratios} and other metrics such as Recall@100 and MAP@100 in Appendix \ref{sssec:main_results_recall_map}, which show the same trend as the results in this section.

\noindent \textbf{ERA enables effective asymmetric retrieval without updating the backbone models or re-indexing the corpus.}
As shown in the table, asymmetric ERA retrieval consistently outperforms zero-shot retrieval and symmetric ERA retrieval across all document embedders, demonstrating the effectiveness of ERA in enabling asymmetric retrieval even when the query and document embedders are from different model families. 
Notably, the performance improvement is more significant when the document embedder is weaker, i.e., more than \textbf{12\% nDCG@10 improvement for OpenAI-small and BGE-M3}, which suggests that ERA can effectively bridge the representation gap between a stronger query embedder and a weaker document embedder, allowing for more effective retrieval adaptation without the need for re-indexing the corpus.   

\noindent \textbf{ERA's alignment stage is a key driver of its performance, and supervised adaptation further enhances the results.}
Comparing ERA w/o adapt. with zero-shot retrieval, we see that the alignment stage alone already provides consistent improvements across models and domains, highlighting the value of bridging the representation gap between the query and document embedders. 
This is especially evident for asymmetric ERA, where alignment substantially improves retrieval even before supervised adaptation. Full ERA then adds further gains on top of this aligned starting point, indicating that supervised adaptation captures semantic distinctions between queries and documents that alignment alone does not model. 
We conjecture that part of the alignment-only gain comes from training on unlabeled documents from the target corpus itself, which adapts the query-side representation to the corpus geometry and makes retrieval more compatible even before using labeled pairs.
}

\begin{figure*}[t]
   \centering
   \begin{subfigure}[b]{0.32\linewidth}
      \centering
      \includegraphics[width=.9\linewidth]{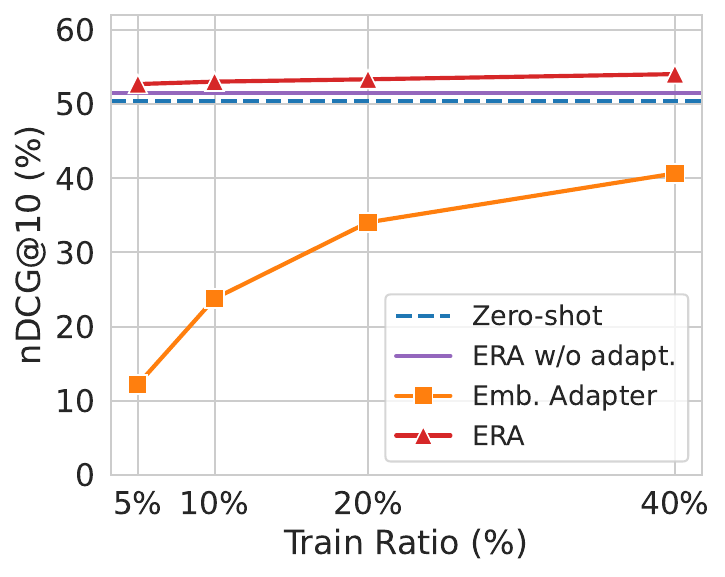}
      \caption{Qwen3-8B -- Qwen3-0.6B}
      \label{fg:asym_train_ratio_qwen3}
   \end{subfigure}
   \hfill
   \begin{subfigure}[b]{0.32\linewidth}
      \centering
      \includegraphics[width=.9\linewidth]{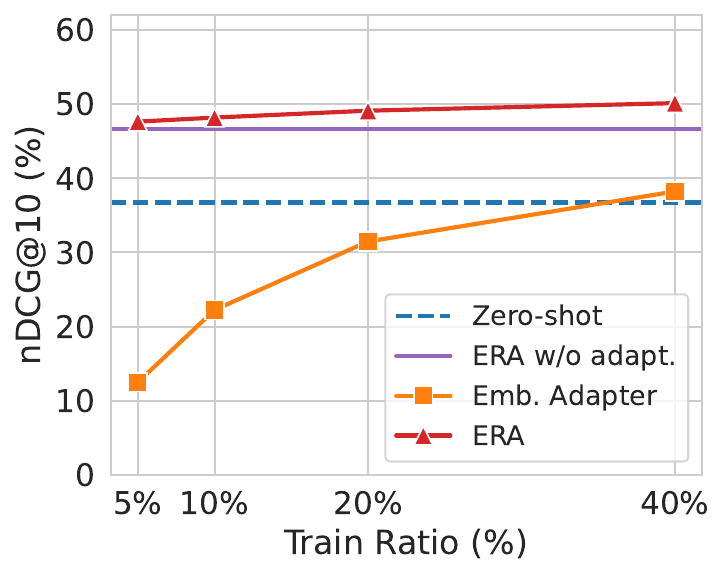}
      \caption{Qwen3-8B -- OpenAI-small}
      \label{fg:asym_train_ratio_openai}
   \end{subfigure}
   \hfill
   \begin{subfigure}[b]{0.32\linewidth}
      \centering
      \includegraphics[width=.9\linewidth]{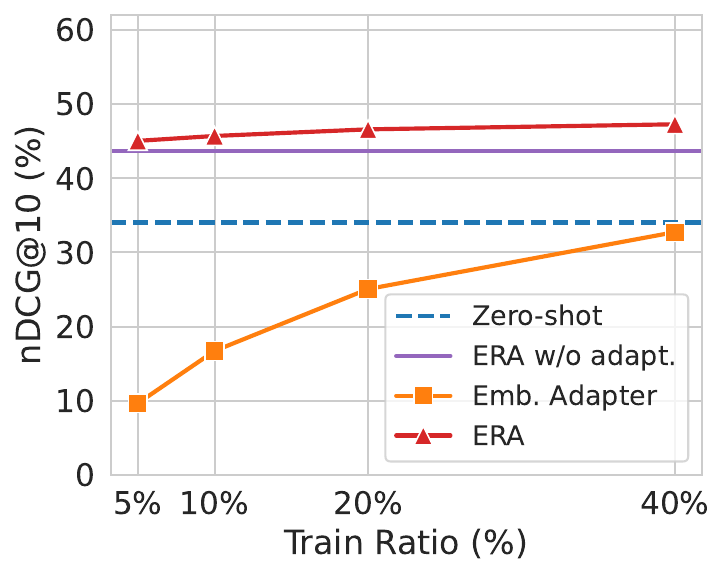}  
      \caption{Qwen3-8B -- BGE-M3}
      \label{fg:asym_train_ratio_bge}
   \end{subfigure}
   \caption{nDCG@10 of zero-shot, ERA w/o adaptation, Embedding Adapter, and ERA retrieval at varying train ratios, where we use Qwen3-8B as the query embedder.}
   \label{fg:label_efficiency}
\end{figure*}

\begin{figure*}[t]
   \centering
   \begin{subfigure}[b]{0.47\linewidth}
      \centering
      \includegraphics[width=.9\linewidth]{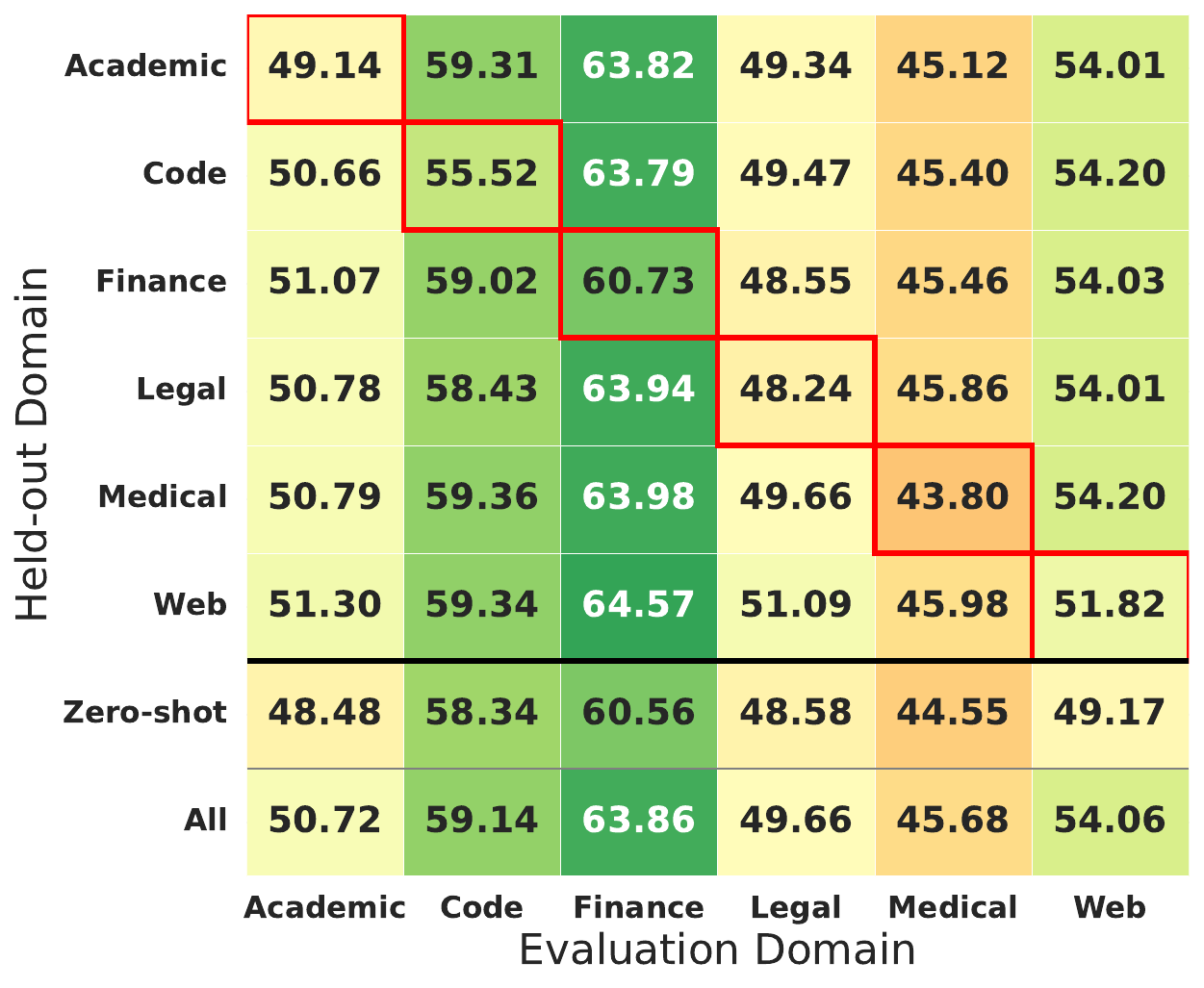}
      \caption{nDCG@10 comparison between zero-shot, ERA trained on all domains, and ERA trained on all domains except the target domain.}
      \label{fg:out-of-domain}
   \end{subfigure}
   \hfill
   \begin{subfigure}[b]{0.49\linewidth}
      \centering
      \includegraphics[width=.9\linewidth]{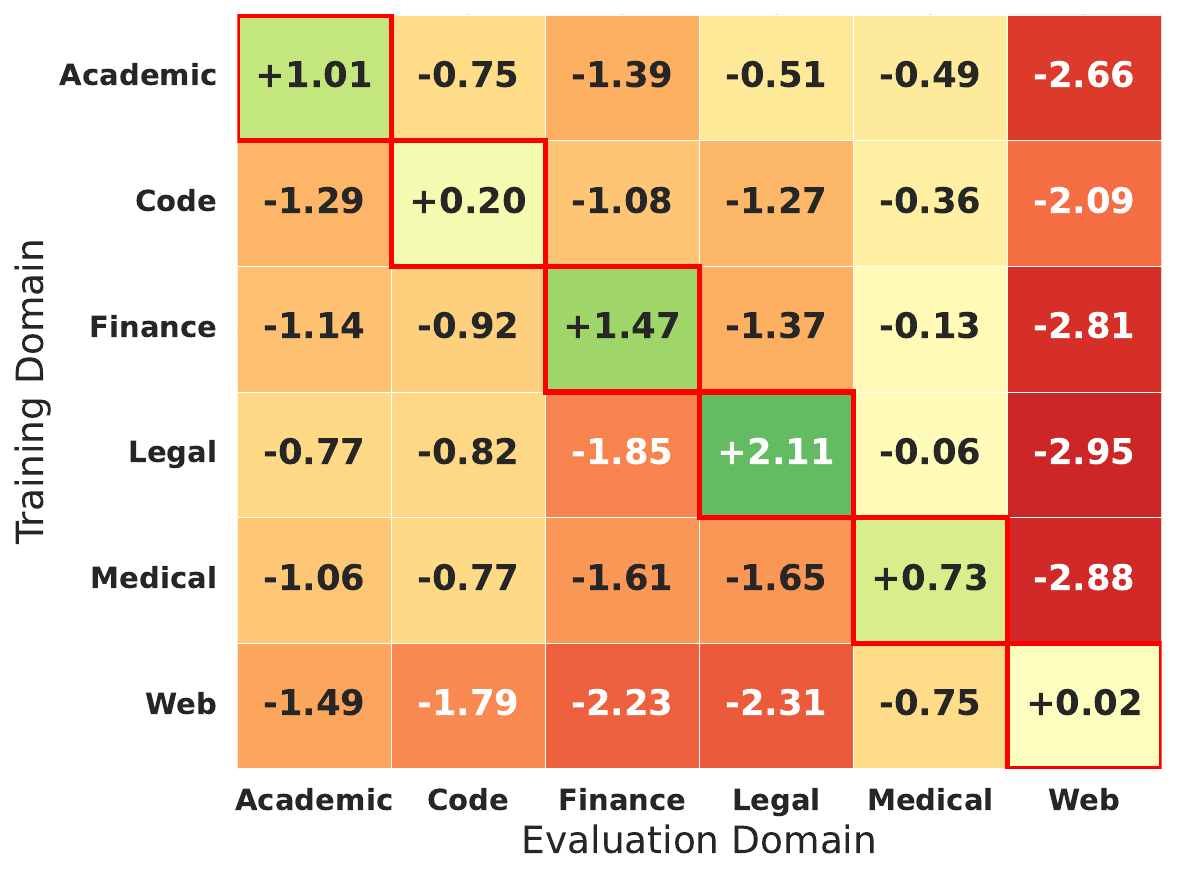}
      \caption{nDCG@10 difference (ERA trained on only the target domain $-$ ERA trained on all domains); positive values indicate that training on only the target domain yields better performance.}
      \label{fg:domain_specific}
   \end{subfigure}
   \caption{Domain generality analysis using Qwen3-8B as the query embedder and Qwen3-0.6B as the document embedder, trained on 20\% of labeled query-document pairs.}
   \label{fg:domain_generality}
\end{figure*}

\subsection{Label Efficiency (RQ2)}
\label{ssec:label_efficiency}
To evaluate the label efficiency of ERA, we conduct experiments with varying percentages of labeled query-document pairs for training, ranging from 5\% to 40\%. We use Qwen3-8B as the query embedder and Qwen3-0.6B, OpenAI-small, and BGE-M3 as the document embedders for ERA and Embedding Adapter, while zero-shot retrieval uses the same embedder for both query and document encoding.
Figure \ref{fg:label_efficiency} shows the results of asymmetric ERA and baselines across different label resource settings. Since zero-shot and ERA w/o adaptation do not use any labeled examples, their performance remains constant thus is shown as horizontal lines.

\noindent \textbf{ERA consistently improves retrieval performance across low-label resource settings.}
ERA consistently outperforms zero-shot retrieval and Embedding Adapter across all training ratios. Even with only 5\% of labeled query-document pairs, i.e., 500 queries in total, ERA still yields clear gains over zero-shot retrieval.

\noindent \textbf{Alignment matters most in the lowest-label settings.}
Across all training ratios, ERA w/o adaptation remains above zero-shot retrieval, and the large gap between ERA and Embedding Adapter in the 5\% and 10\% settings indicates that alignment is what makes the supervised stage effective in the low-label regime.

\subsection{Domain Generality (RQ3)}
\label{ssec:domain_general}
To evaluate the domain generality of ERA, we conduct two sets of experiments with Qwen3-8B as the query embedder and Qwen3-0.6B as the document embedder: \noindent \textbf{1) Out-of-Domain Experiment}, where we train ERA on all domains except the target domain, and \textbf{2) Domain-specific Experiment}, where we train ERA using only the target domain.
We use $20\%$ of labeled query-document pairs for training in both sets of experiments. 
The results of other train ratios are provided in Appendix \ref{app:ssec:domain_generality_train_ratios}, which show the same trends as the results with 20\% of labeled query-document pairs.

\noindent \textbf{(1. Out-of-Domain) ERA is robust against out-of-domain tasks.}
As shown in Figure \ref{fg:out-of-domain}, while ERA trained on all domains except the target domain shows comparable or slightly lower performance than zero-shot retrieval for the target domain (see the diagonal elements in the top six rows), it improves the performance in other domains over zero-shot. 
This indicates that training on other domains does not significantly degrade the performance on out-of-domain tasks. 
We conjecture that ERA's linear transformation is simple enough to avoid overfitting to the training domains, which allows it to maintain its generalization ability to out-of-domain tasks. 

\noindent \textbf{(2. Domain-specific) A general adapter trained across all domains can perform comparably with a domain-specific adapter while excelling in other domains.}
As shown in Figure \ref{fg:domain_specific}, while the performance of domain-specific adapters trained on only the target domain can be slightly better than the general adapter trained on each domain where they are trained, the performance difference is generally small across all domains.
This encourages the use of a single general adapter for retrieval adaptation across different domains, avoiding the need for domain-specific adaptations and switching between different adapters for different domains, which can increase the complexity of the system, introduce maintenance overhead, and scale poorly with the number of domains.

\subsection{Ablation Study (RQ4)}
\label{ssec:design_choices}
We conduct ablation studies on the negative sampling strategy, the number of negative samples\cameraready{, and the adapter complexity}. 
The hyperparameter analysis on weight decay and learning rate is provided in Appendix \ref{sssec:hp_search_results}.

\begin{figure}
   \centering
   \includegraphics[width=.95\linewidth]{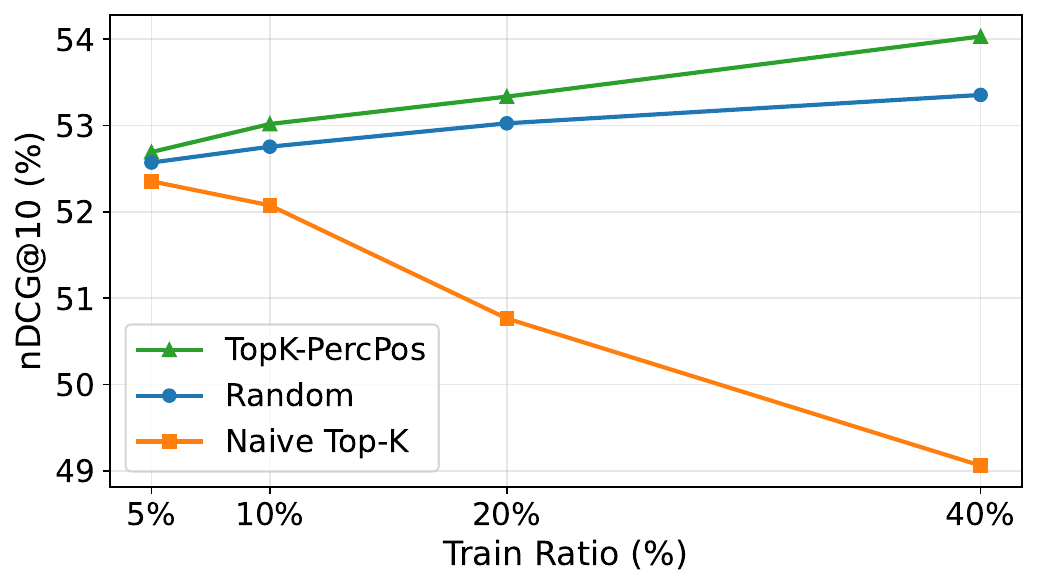}
   \caption{Ablation on negative sampling strategy with varying train ratios.}
   \label{fg:neg_strategy}
\end{figure}

\subsubsection{Negative Sampling Strategy}
\label{sssec:neg_sampling_strategy}
We use Qwen3-8B as the query embedder and Qwen3-0.6B as the document embedder. We compare TopK-PercPos, naive top-k negative sampling, and random negative sampling. Naive top-k selects the top-k most similar documents to the query as negatives based on the base embedder's similarity scores, while random sampling selects $k$ documents uniformly from the corpus.

\noindent \textbf{TopK-PercPos negative sampling is more effective than naive top-k and random negative sampling.} 
As shown in Figure \ref{fg:neg_strategy}, TopK-PercPos consistently outperforms both naive top-k and random negative sampling across different train ratios, demonstrating the importance of selecting hard negatives based on retrieval scores after the alignment stage for effective supervised adaptation. 
As reported by \citet{gabriel2025hardnegative}, naive top-k negatives can degrade performance by selecting false negatives, i.e., relevant documents mislabeled as negatives. 

\begin{table}
  \centering
  \scalebox{0.8}{
  \begin{tabular}{lccccc}
    \toprule
    \# Negatives & 5 & 10 & 50 & 100 & 200 \\
    \midrule
    nDCG@10 (\%) & \textbf{53.02} & 52.80 & 52.89 & 52.65 & 52.66 \\
    \bottomrule
  \end{tabular}
  }
  \caption{Number of negative samples ablation. We set the train ratio to 10\%.}
  \label{tab:ablation_num_negatives}
\end{table}

\subsubsection{Number of Negative Samples}
\label{sssec:num_negatives}
We evaluate the impact of the number of negative samples on the performance of ERA under TopK-PercPos negative sampling strategy. We vary the number of negative samples from 5 to 200 and evaluate the nDCG@10 with a train ratio of 10\%.

\noindent \textbf{ERA is robust against the number of negative samples.}
As shown in Table \ref{tab:ablation_num_negatives}, the performance of ERA is relatively stable across different numbers of negative samples, with the best performance achieved with 5 negative samples.
This suggests that ERA can be effective even with a small number of negative samples, which can reduce the computational cost of training while still maintaining strong retrieval performance. 

\cameraready{
\subsubsection{Adapter Complexity}
We evaluate the impact of adapter complexity on the performance of ERA by comparing a linear adapter with a two-layer MLP adapter which has 1024 hidden units and ReLU activation. The other setup and hyperparameter search space are the same as those described in Section \ref{ssec:setup}. Through the hyperparameter search, we set the weight decay and learning rate to 0.01 and 5e-5, respectively. 

Table \ref{tab:mlp_vs_linear} shows that ERA with MLP obtains similar retrieval performance in all metrics. To provide a more detailed comparison, we investigate how many tasks in MAIR (126 tasks) MLP outperforms Linear. The nDCG@10 results show MLP wins 41 tasks while Linear wins 84 tasks (ties in one task). We conjecture that a linear weight is sufficiently capable for ERA and does not overfit to a small set of training samples due to its simplicity. This observation corresponds to the findings in \cite{chromadb2024embedding}.

\begin{table}[t]
  \centering
  \resizebox{\linewidth}{!}{
  \begin{tabular}{lrrr}
    \toprule
    Method & nDCG@10 & Recall@100 & MAP@100 \\
    \midrule
    MLP ERA w/o adapt. & 50.33 & 70.79 & 39.74 \\
    MLP ERA & 51.84 & 72.49 & 41.27 \\
    \textbf{Linear ERA} & \textbf{53.02} & \textbf{72.96} & \textbf{42.39} \\
    \bottomrule
  \end{tabular}
  }
  \caption{Adapter complexity analysis (\%). We set
the train ratio to 10\%, a query encoder to Qwen3-8B, and a document encoder to Qwen3-0.6B. }
  \label{tab:mlp_vs_linear}
\end{table}

}

\section{Conclusion}
\label{sec:conclusion}
We propose Efficient Retrieval Adapter (ERA), a novel framework for training retrieval adapters in a label-efficient manner to address the challenges of retrieval adaptation under query-document complexity asymmetry. 
ERA employs a two-stage training process, \textit{self-supervised alignment training} and \textit{supervised adaptation}, to effectively bridge the representation gap between different embedding models and the semantic gap between complex queries and simple documents.
Our experiments demonstrate that ERA efficiently and effectively improves retrieval performance across various models, domains, and label resource settings, and enables asymmetric retrieval with different query and document embedders without the need for re-indexing the corpus.

\section*{Limitations}
\seiji{
While ERA improves retrieval performance without updating backbone embedders or re-indexing the document corpus, it has several limitations. First, ERA uses a linear adapter, and we do not investigate whether non-linear or deeper adapters further improve cross-model alignment. Second, although ERA avoids full-corpus re-indexing, the alignment stage still requires embedding a small sampled set of unlabeled documents with both query and document embedders. As we mentioned in Section \ref{sec:exp}, we sample only 1,000 documents for each training task, which is a substantially smaller number than the size of the full corpus. Third, while we evaluate ERA on a wide range of retrieval tasks across different domains, we do not evaluate it on other retrieval settings such as multi-modal retrieval and cross-lingual retrieval, which can be interesting directions for future work.
Lastly, while there are many potential strategies for the retrieval performance improvement under the setting of frozen backbone embedders and fixed document index such as sparse and dense fusion, re-ranking, document-side adaptation, and few-shot prompting, we focus on the retrieval adapter approach in this work. We leave the exploration of the combination of retrieval adapters with other strategies for future work because they are complementary to each other and can potentially lead to further retrieval performance improvement rather than a direct comparison between them.
}

\section*{Ethical Considerations}
Our work focuses on improving retrieval performance by training retrieval adapters in a label-efficient manner.
While we do not foresee any direct ethical concerns with our method, it is important to consider the potential downstream applications of retrieval systems that utilize ERA, such as information retrieval, question answering, and recommendation systems. 
These applications may have ethical implications related to user privacy, bias in retrieved results, and the potential for misuse in spreading misinformation. 
Therefore, it is crucial for practitioners to carefully evaluate and mitigate any ethical concerns when deploying retrieval systems that incorporate ERA, and to ensure that the benefits of improved retrieval performance are balanced with responsible and ethical use.

We made use of AI tools such as ChatGPT and Copilot to support coding and refining this paper, but all content was carefully reviewed and edited by us to ensure it adheres to our standards and aligns with our research objectives.

\bibliography{custom}

@inproceedings{
thakur2021beir,
title={{BEIR}: A Heterogeneous Benchmark for Zero-shot Evaluation of Information Retrieval Models},
author={Nandan Thakur and Nils Reimers and Andreas R{\"u}ckl{\'e} and Abhishek Srivastava and Iryna Gurevych},
booktitle={Thirty-fifth Conference on Neural Information Processing Systems Datasets and Benchmarks Track (Round 2)},
year={2021},
url={https://openreview.net/forum?id=wCu6T5xFjeJ}
}

@inproceedings{devlin-etal-2019-bert,
    title = "{BERT}: Pre-training of Deep Bidirectional Transformers for Language Understanding",
    author = "Devlin, Jacob  and
      Chang, Ming-Wei  and
      Lee, Kenton  and
      Toutanova, Kristina",
    editor = "Burstein, Jill  and
      Doran, Christy  and
      Solorio, Thamar",
    booktitle = "Proceedings of the 2019 Conference of the North {A}merican Chapter of the Association for Computational Linguistics: Human Language Technologies, Volume 1 (Long and Short Papers)",
    month = jun,
    year = "2019",
    address = "Minneapolis, Minnesota",
    publisher = "Association for Computational Linguistics",
    url = "https://aclanthology.org/N19-1423/",
    doi = "10.18653/v1/N19-1423",
    pages = "4171--4186"
}

@inproceedings{
loshchilov2018decoupled,
title={Decoupled Weight Decay Regularization},
author={Ilya Loshchilov and Frank Hutter},
booktitle={International Conference on Learning Representations},
year={2019},
url={https://openreview.net/forum?id=Bkg6RiCqY7},
}

@article{oord2018representation,
  title={Representation learning with contrastive predictive coding},
  author={Oord, Aaron van den and Li, Yazhe and Vinyals, Oriol},
  url={https://arxiv.org/abs/1807.03748},
  journal={arXiv preprint arXiv:1807.03748},
  year={2018}
}

@inproceedings{taguchi-etal-2025-efficient,
    title = "Efficient Context Selection for Long-Context {QA}: No Tuning, No Iteration, Just Adaptive{-}$k$",
    author = "Taguchi, Chihiro  and
      Maekawa, Seiji  and
      Bhutani, Nikita",
    editor = "Christodoulopoulos, Christos  and
      Chakraborty, Tanmoy  and
      Rose, Carolyn  and
      Peng, Violet",
    booktitle = "Proceedings of the 2025 Conference on Empirical Methods in Natural Language Processing",
    month = nov,
    year = "2025",
    address = "Suzhou, China",
    publisher = "Association for Computational Linguistics",
    url = "https://aclanthology.org/2025.emnlp-main.1017/",
    doi = "10.18653/v1/2025.emnlp-main.1017",
    pages = "20105--20130",
    ISBN = "979-8-89176-332-6"
}

@inproceedings{maekawa-etal-2024-retrieval,
    title = "Retrieval Helps or Hurts? A Deeper Dive into the Efficacy of Retrieval Augmentation to Language Models",
    author = "Maekawa, Seiji  and
      Iso, Hayate  and
      Gurajada, Sairam  and
      Bhutani, Nikita",
    editor = "Duh, Kevin  and
      Gomez, Helena  and
      Bethard, Steven",
    booktitle = "Proceedings of the 2024 Conference of the North American Chapter of the Association for Computational Linguistics: Human Language Technologies (Volume 1: Long Papers)",
    month = jun,
    year = "2024",
    address = "Mexico City, Mexico",
    publisher = "Association for Computational Linguistics",
    url = "https://aclanthology.org/2024.naacl-long.308/",
    doi = "10.18653/v1/2024.naacl-long.308",
    pages = "5506--5521"
}

@inproceedings{huh2024position,
title={Position: The Platonic Representation Hypothesis},
author={Minyoung Huh and Brian Cheung and Tongzhou Wang and Phillip Isola},
booktitle={Forty-first International Conference on Machine Learning},
year={2024},
url={https://openreview.net/forum?id=BH8TYy0r6u}
}

@inproceedings{schlatt2025distillm,
author = {Schlatt, Ferdinand and Fr\"{o}be, Maik and Scells, Harrisen and Zhuang, Shengyao and Koopman, Bevan and Zuccon, Guido and Stein, Benno and Potthast, Martin and Hagen, Matthias},
title = {Rank-DistiLLM: Closing the Effectiveness Gap Between Cross-Encoders and LLMs for Passage Re-ranking},
year = {2025},
isbn = {978-3-031-88713-0},
publisher = {Springer-Verlag},
address = {Berlin, Heidelberg},
url = {https://doi.org/10.1007/978-3-031-88714-7_31},
doi = {10.1007/978-3-031-88714-7_31},
booktitle = {Advances in Information Retrieval: 47th European Conference on Information Retrieval, ECIR 2025, Lucca, Italy, April 6–10, 2025, Proceedings, Part III},
pages = {323–334},
numpages = {12},
location = {Lucca, Italy}
}

@inproceedings{vejendla-2025-drift,
    title = "Drift-Adapter: A Practical Approach to Near Zero-Downtime Embedding Model Upgrades in Vector Databases",
    author = "Vejendla, Harshil",
    editor = "Christodoulopoulos, Christos  and
      Chakraborty, Tanmoy  and
      Rose, Carolyn  and
      Peng, Violet",
    booktitle = "Proceedings of the 2025 Conference on Empirical Methods in Natural Language Processing",
    month = nov,
    year = "2025",
    address = "Suzhou, China",
    publisher = "Association for Computational Linguistics",
    url = "https://aclanthology.org/2025.emnlp-main.805/",
    doi = "10.18653/v1/2025.emnlp-main.805",
    pages = "15938--15949",
    ISBN = "979-8-89176-332-6"
}

@inproceedings{karpukhin-etal-2020-dense,
    title = "Dense Passage Retrieval for Open-Domain Question Answering",
    author = "Karpukhin, Vladimir  and
      Oguz, Barlas  and
      Min, Sewon  and
      Lewis, Patrick  and
      Wu, Ledell  and
      Edunov, Sergey  and
      Chen, Danqi  and
      Yih, Wen-tau",
    editor = "Webber, Bonnie  and
      Cohn, Trevor  and
      He, Yulan  and
      Liu, Yang",
    booktitle = "Proceedings of the 2020 Conference on Empirical Methods in Natural Language Processing (EMNLP)",
    month = nov,
    year = "2020",
    address = "Online",
    publisher = "Association for Computational Linguistics",
    url = "https://aclanthology.org/2020.emnlp-main.550/",
    doi = "10.18653/v1/2020.emnlp-main.550",
    pages = "6769--6781"
}

@inproceedings{gabriel2025hardnegative,
author = {Moreira, Gabriel de Souza P. and Osmulski, Radek and Xu, Mengyao and Ak, Ronay and Schifferer, Benedikt and Oldridge, Even},
title = {Improving Text Embedding Models with Positive-aware Hard-negative Mining},
year = {2025},
isbn = {9798400720406},
publisher = {Association for Computing Machinery},
address = {New York, NY, USA},
url = {https://doi.org/10.1145/3746252.3761254},
doi = {10.1145/3746252.3761254},
booktitle = {Proceedings of the 34th ACM International Conference on Information and Knowledge Management},
pages = {2169–2178},
numpages = {10},
location = {Seoul, Republic of Korea},
series = {CIKM '25}
}

@inproceedings{petroni-etal-2021-kilt,
    title = "{KILT}: a Benchmark for Knowledge Intensive Language Tasks",
    author = {Petroni, Fabio  and
      Piktus, Aleksandra  and
      Fan, Angela  and
      Lewis, Patrick  and
      Yazdani, Majid  and
      De Cao, Nicola  and
      Thorne, James  and
      Jernite, Yacine  and
      Karpukhin, Vladimir  and
      Maillard, Jean  and
      Plachouras, Vassilis  and
      Rockt{\"a}schel, Tim  and
      Riedel, Sebastian},
    editor = "Toutanova, Kristina  and
      Rumshisky, Anna  and
      Zettlemoyer, Luke  and
      Hakkani-Tur, Dilek  and
      Beltagy, Iz  and
      Bethard, Steven  and
      Cotterell, Ryan  and
      Chakraborty, Tanmoy  and
      Zhou, Yichao",
    booktitle = "Proceedings of the 2021 Conference of the North American Chapter of the Association for Computational Linguistics: Human Language Technologies",
    month = jun,
    year = "2021",
    address = "Online",
    publisher = "Association for Computational Linguistics",
    url = "https://aclanthology.org/2021.naacl-main.200/",
    doi = "10.18653/v1/2021.naacl-main.200",
    pages = "2523--2544"
}

@article{zhang2025qwen3embedding,
  title={Qwen3 embedding: Advancing text embedding and reranking through foundation models},
  author={Zhang, Yanzhao and Li, Mingxin and Long, Dingkun and Zhang, Xin and Lin, Huan and Yang, Baosong and Xie, Pengjun and Yang, An and Liu, Dayiheng and Lin, Junyang and others},
  journal={arXiv preprint arXiv:2506.05176},
  url={https://arxiv.org/abs/2506.05176},
  year={2025}
}

@inproceedings{
lee2025nvembed,
title={{NV}-Embed: Improved Techniques for Training {LLM}s as Generalist Embedding Models},
author={Chankyu Lee and Rajarshi Roy and Mengyao Xu and Jonathan Raiman and Mohammad Shoeybi and Bryan Catanzaro and Wei Ping},
booktitle={The Thirteenth International Conference on Learning Representations},
year={2025},
url={https://openreview.net/forum?id=lgsyLSsDRe}
}

@inproceedings{
chen2025transferring,
title={Transferring Linear Features Across Language Models With Model Stitching},
author={Alan Chen and Jack Merullo and Alessandro Stolfo and Ellie Pavlick},
booktitle={The Thirty-ninth Annual Conference on Neural Information Processing Systems},
year={2025},
url={https://openreview.net/forum?id=Qvvy0X63Fv}
}

@inproceedings{
jha2025harnessing,
title={Harnessing the Universal Geometry of Embeddings},
author={Rishi Dev Jha and Collin Zhang and Vitaly Shmatikov and John Xavier Morris},
booktitle={The Thirty-ninth Annual Conference on Neural Information Processing Systems},
year={2025},
url={https://openreview.net/forum?id=jiCLUPq5xv}
}

@inproceedings{elsen2025nevir,
author = {van den Elsen, Coen and Barkhof, Francien and Nijdam, Thijmen and Lupart, Simon and Aliannejadi, Mohammad},
title = {Reproducing NevIR: Negation in Neural Information Retrieval},
year = {2025},
isbn = {9798400715921},
publisher = {Association for Computing Machinery},
address = {New York, NY, USA},
url = {https://doi.org/10.1145/3726302.3730294},
doi = {10.1145/3726302.3730294},
booktitle = {Proceedings of the 48th International ACM SIGIR Conference on Research and Development in Information Retrieval},
pages = {3346–3356},
numpages = {11},
location = {Padua, Italy},
series = {SIGIR '25}
}

@inproceedings{weller-etal-2025-followir,
    title = "{F}ollow{IR}: Evaluating and Teaching Information Retrieval Models to Follow Instructions",
    author = "Weller, Orion  and
      Chang, Benjamin  and
      MacAvaney, Sean  and
      Lo, Kyle  and
      Cohan, Arman  and
      Van Durme, Benjamin  and
      Lawrie, Dawn  and
      Soldaini, Luca",
    editor = "Chiruzzo, Luis  and
      Ritter, Alan  and
      Wang, Lu",
    booktitle = "Proceedings of the 2025 Conference of the Nations of the Americas Chapter of the Association for Computational Linguistics: Human Language Technologies (Volume 1: Long Papers)",
    month = apr,
    year = "2025",
    address = "Albuquerque, New Mexico",
    publisher = "Association for Computational Linguistics",
    url = "https://aclanthology.org/2025.naacl-long.597/",
    doi = "10.18653/v1/2025.naacl-long.597",
    pages = "11926--11942",
    ISBN = "979-8-89176-189-6"
}

@inproceedings{sun-etal-2024-mair,
    title = "{MAIR}: A Massive Benchmark for Evaluating Instructed Retrieval",
    author = "Sun, Weiwei  and
      Shi, Zhengliang  and
      Long, Wu Jiu  and
      Yan, Lingyong  and
      Ma, Xinyu  and
      Liu, Yiding  and
      Cao, Min  and
      Yin, Dawei  and
      Ren, Zhaochun",
    editor = "Al-Onaizan, Yaser  and
      Bansal, Mohit  and
      Chen, Yun-Nung",
    booktitle = "Proceedings of the 2024 Conference on Empirical Methods in Natural Language Processing",
    month = nov,
    year = "2024",
    address = "Miami, Florida, USA",
    publisher = "Association for Computational Linguistics",
    url = "https://aclanthology.org/2024.emnlp-main.778/",
    doi = "10.18653/v1/2024.emnlp-main.778",
    pages = "14044--14067"
}

@inproceedings{kim-baek-2025-syntriever,
    title = "Syntriever: How to Train Your Retriever with Synthetic Data from {LLM}s",
    author = "Kim, Minsang  and
      Baek, Seung Jun",
    editor = "Chiruzzo, Luis  and
      Ritter, Alan  and
      Wang, Lu",
    booktitle = "Findings of the Association for Computational Linguistics: NAACL 2025",
    month = apr,
    year = "2025",
    address = "Albuquerque, New Mexico",
    publisher = "Association for Computational Linguistics",
    url = "https://aclanthology.org/2025.findings-naacl.136/",
    doi = "10.18653/v1/2025.findings-naacl.136",
    pages = "2523--2539",
    ISBN = "979-8-89176-195-7"
}

@inproceedings{chen-etal-2024-m3,
    title = "{M}3-Embedding: Multi-Linguality, Multi-Functionality, Multi-Granularity Text Embeddings Through Self-Knowledge Distillation",
    author = "Chen, Jianlyu  and
      Xiao, Shitao  and
      Zhang, Peitian  and
      Luo, Kun  and
      Lian, Defu  and
      Liu, Zheng",
    editor = "Ku, Lun-Wei  and
      Martins, Andre  and
      Srikumar, Vivek",
    booktitle = "Findings of the Association for Computational Linguistics: ACL 2024",
    month = aug,
    year = "2024",
    address = "Bangkok, Thailand",
    publisher = "Association for Computational Linguistics",
    url = "https://aclanthology.org/2024.findings-acl.137/",
    doi = "10.18653/v1/2024.findings-acl.137",
    pages = "2318--2335"
}

@inproceedings{
dai2023promptagator,
title={Promptagator: Few-shot Dense Retrieval From 8 Examples},
author={Zhuyun Dai and Vincent Y Zhao and Ji Ma and Yi Luan and Jianmo Ni and Jing Lu and Anton Bakalov and Kelvin Guu and Keith Hall and Ming-Wei Chang},
booktitle={The Eleventh International Conference on Learning Representations },
year={2023},
url={https://openreview.net/forum?id=gmL46YMpu2J}
}

@inproceedings{yoon-etal-2024-search,
    title = "Search-Adaptor: Embedding Customization for Information Retrieval",
    author = "Yoon, Jinsung  and
      Chen, Yanfei  and
      Arik, Sercan  and
      Pfister, Tomas",
    editor = "Ku, Lun-Wei  and
      Martins, Andre  and
      Srikumar, Vivek",
    booktitle = "Proceedings of the 62nd Annual Meeting of the Association for Computational Linguistics (Volume 1: Long Papers)",
    month = aug,
    year = "2024",
    address = "Bangkok, Thailand",
    publisher = "Association for Computational Linguistics",
    url = "https://aclanthology.org/2024.acl-long.661/",
    doi = "10.18653/v1/2024.acl-long.661",
    pages = "12230--12247"
}

@article{openai2024embedding,
    author = {OpenAI},
    title = {New embedding models and API updates},
    url = {https://openai.com/index/new-embedding-models-and-api-updates/},
    journal = {Technical report},
    year = {2024}
}

@article{chromadb2024embedding,
    author = {Sanjeev, Suvansh and Troynikov, Anton},
    title = {Embedding Adapters},
    url = {https://research.trychroma.com/embedding-adapters},
    journal = {Chroma Technical Report},
    year = {2024}
}

@inproceedings{reimers-gurevych-2019-sentence,
    title = "Sentence-{BERT}: Sentence Embeddings using {S}iamese {BERT}-Networks",
    author = "Reimers, Nils  and
      Gurevych, Iryna",
    editor = "Inui, Kentaro  and
      Jiang, Jing  and
      Ng, Vincent  and
      Wan, Xiaojun",
    booktitle = "Proceedings of the 2019 Conference on Empirical Methods in Natural Language Processing and the 9th International Joint Conference on Natural Language Processing (EMNLP-IJCNLP)",
    month = nov,
    year = "2019",
    address = "Hong Kong, China",
    publisher = "Association for Computational Linguistics",
    url = "https://aclanthology.org/D19-1410/",
    doi = "10.18653/v1/D19-1410",
    pages = "3982--3992"
}

\appendix
\section{Embedding Models}
\label{app:embedders}

We summarize the details of the models used in our experiments in Table \ref{tab:models}. 
All models are accessed through HuggingFace except for OpenAI-small, which is accessed through the OpenAI API.
We adhere their intended usage and licenses for all models since the models are used for retrieval purposes in our experiments. 

\footnotetext[1]{\url{https://openai.com/policies/services-agreement/} [Accessed: May 22, 2026]}

\begin{table*}[h]
  \centering
  \small
  \resizebox{0.9\linewidth}{!}{
  \begin{tabular}{lrrrll}
    \toprule
    \textbf{Model}
      & \textbf{Size}
      & \textbf{Context}
      & \textbf{Dimension}
      & \textbf{HuggingFace / API}
      & \textbf{License} \\
    \midrule
    Qwen3-8B \citep{zhang2025qwen3embedding}
      & 8B & 32k & 4096
      & \texttt{Qwen/Qwen3-Embedding-8B}
      & Apache license 2.0 \\
    Qwen3-0.6B \citep{zhang2025qwen3embedding}
      & 0.6B & 32k & 1024
      & \texttt{Qwen/Qwen3-Embedding-0.6B}
      & Apache license 2.0 \\
    OpenAI-small \citep{openai2024embedding}
      & - & 8k & 1536
      & \texttt{text-embedding-3-small}
      & OpenAI Service Terms\footnotemark[1] \\
    BGE-M3 \citep{chen-etal-2024-m3}
      & 0.6B & 8k & 1024
      & \texttt{BAAI/bge-m3}
      & MIT \\
    \bottomrule
  \end{tabular}
  }
  \caption{Models used in experiments. Dimensions denote the output embedding sizes. Model sizes are not publicly disclosed (-).}
  \label{tab:models}
\end{table*}

\section{Experiment Details}
\label{app:exp_details}
We provide additional details of our experiments in this section, including the hyperparameters used for training ERA and baselines, the negative sampling strategy, and the evaluation metrics.
\subsection{Training Details}
\label{sssec:training_hyperparameters}
We use the AdamW optimizer \citep{loshchilov2018decoupled} in both stages. 
For the label-based training stage, we apply linear warmup for the first 10\% of training steps. When using the InfoNCE loss, we set the temperature to $0.05$.

For the adaptation stage, we test different learning rates (1e-3, 5e-4, 1e-4, 5e-5, 1e-5) and weight decays (1e-2, 1e-3, 1e-4) with Qwen3-8B as a query embedder and Qwen3-0.6B as a document embedder, trained on 10\% of labeled query-document pairs. 
While we find that performance is generally stable across these hyperparameters, we use for all experiments the hyperparameter set described in Section \ref{ssec:setup}, which yields the best retrieval evaluation performance. 
We also tune the hyperparameters for the Embedding Adapter and use the best set, 1e-3 for the learning rate and 1e-4 for weight decay, which we use for all experiments.
We show the hyperparameter search results in the Appendix \ref{sssec:hp_search_results}.

\subsection{Token Cost Analysis}
\label{app:ssec:token_cost_analysis}
The alignment stage requires encoding only a small subset of the MAIR corpus rather than the full document collection. We compute token counts using the \texttt{cl100k\_base} tokenizer. Across MAIR, the full corpus contains 8,958,270 documents with 2,549,625,873 tokens in total, whereas the alignment sample used in our experiments contains 114,858 documents with 132,526,211 tokens. This corresponds to 1.28\% of the documents and 5.20\% of the total corpus tokens. Therefore, the additional cost of the alignment stage is substantially small relative to full-corpus processing.

\section{Additional Results}
\label{app:additional_results}

\subsection{Main Results with Varying Train Ratios}
\label{sssec:main_results_varying_train_ratios}

We show the nDCG@10 results of ERA and baselines on the MAIR benchmark using varying percentages of labeled query-document pairs for training, including 5\%, 10\%, and 40\% in Tables \ref{tab:main_results_5percent}, \ref{tab:main_results_10percent}, and \ref{tab:main_results_40percent}, respectively.
We observe the same trend that ERA consistently improves retrieval performance across all models in most domains, compared with zero-shot retrieval and Embedding Adapter, demonstrating the effectiveness of ERA in improving retrieval performance under different label resource settings.

\begin{table*}[!t]
  \centering
  \small
  \scalebox{0.85}{
  \begin{tabular}{llrrrrrrr}
    \toprule
    Base model & Variant & \textbf{Avg} & Academic & Code & Finance & Legal & Medical & Web \\
    \midrule
    \multirow{6}{*}{BGE-M3} & Zero-shot & 34.03 & 34.81 & 34.85 & 52.44 & 37.01 & 25.25 & 33.28 \\
     & Emb.\ Adapter & 7.48 & 6.59 & 14.36 & 8.92 & 6.65 & 4.04 & 6.60 \\
     & Sym.\ ERA w/o adapt. & 34.17 & 35.01 & 33.79 & 53.58 & 38.41 & 25.54 & 33.35 \\
     & Sym.\ ERA & 39.80 & 38.75 & 39.62 & 58.45 & 39.60 & 29.65 & 41.02 \\
     & Asym.\ ERA w/o adapt. & 43.69 & 39.15 & 40.16 & \textbf{58.66} & 40.65 & 34.49 & 47.84 \\
     & \textbf{Asym.\ ERA} & \textbf{45.04} & \textbf{41.25} & \textbf{42.48} & 58.47 & \textbf{43.31} & \textbf{35.72} & \textbf{48.65} \\
    \midrule
    \multirow{6}{*}{OpenAI-small} & Zero-shot & 36.74 & 35.62 & 41.35 & 50.36 & 31.77 & 30.64 & 36.67 \\
     & Emb.\ Adapter & 8.36 & 6.47 & 15.62 & 4.85 & 5.82 & 5.75 & 8.46 \\
     & Sym.\ ERA w/o adapt. & 37.19 & 36.32 & 41.01 & 50.13 & 30.79 & 31.09 & 37.71 \\
     & Sym.\ ERA & 41.51 & 40.65 & 44.44 & \textbf{53.63} & 34.21 & 35.38 & 42.65 \\
     & Asym.\ ERA w/o adapt. & 46.63 & 41.69 & 49.48 & 52.68 & 34.72 & 40.40 & 50.86 \\
     & \textbf{Asym.\ ERA} & \textbf{47.63} & \textbf{43.61} & \textbf{50.11} & 52.80 & \textbf{37.65} & \textbf{40.90} & \textbf{51.63} \\
    \midrule
    \multirow{6}{*}{Qwen3-0.6B} & Zero-shot & 50.37 & 48.48 & \textbf{58.34} & 60.56 & 48.58 & 44.55 & 49.17 \\
     & Emb.\ Adapter & 14.83 & 12.21 & 23.04 & 17.22 & 9.31 & 9.53 & 15.49 \\
     & Sym.\ ERA w/o adapt. & 51.12 & 49.28 & 57.28 & 62.17 & \textbf{49.85} & \textbf{45.53} & 50.20 \\
     & Sym.\ ERA & 51.38 & 48.96 & 57.35 & 62.46 & 49.67 & 45.38 & 50.93 \\
     & Asym.\ ERA w/o adapt. & 51.44 & 50.11 & 57.53 & 63.13 & 48.35 & 44.70 & 51.07 \\
     & \textbf{Asym.\ ERA} & \textbf{52.69} & \textbf{50.26} & 57.86 & \textbf{64.09} & 49.62 & 45.36 & \textbf{53.21} \\
    \midrule
    \multirow{4}{*}{Qwen3-8B} & Zero-shot & 57.91 & 56.94 & 61.85 & 69.38 & 57.89 & \textbf{53.75} & 56.65 \\
     & Emb.\ Adapter & 19.77 & 17.72 & 26.67 & 19.25 & 12.84 & 12.31 & 22.18 \\
     & Sym.\ ERA w/o adapt. & 58.51 & \textbf{58.85} & 61.53 & 70.03 & \textbf{58.25} & 53.39 & 57.56 \\
     & Sym.\ ERA & \textbf{58.53} & 58.66 & \textbf{62.03} & \textbf{70.61} & 57.38 & 52.44 & \textbf{57.91} \\
    \bottomrule
  \end{tabular}
  }
  \caption{nDCG@10 (\%) on the MAIR benchmark with 5\% of labeled query-document pairs. }
  \label{tab:main_results_5percent}
\end{table*}

\begin{table*}[!t]
  \centering
  \small
    \scalebox{0.85}{
  \begin{tabular}{llrrrrrrr}
    \toprule
    Base model & Variant & \textbf{Avg} & Academic & Code & Finance & Legal & Medical & Web \\
    \midrule
    \multirow{6}{*}{BGE-M3} & Zero-shot & 34.03 & 34.81 & 34.85 & 52.44 & 37.01 & 25.25 & 33.28 \\
     & Embedding Adapter & 17.37 & 14.68 & 22.16 & 27.52 & 15.27 & 11.67 & 17.51 \\
     & Sym.\ ERA w/o adapt. & 34.17 & 35.01 & 33.79 & 53.58 & 38.41 & 25.54 & 33.35 \\
     & Sym.\ ERA & 40.77 & 39.57 & 38.94 & \textbf{59.36} & 40.47 & 31.15 & 42.43 \\
     & Asym.\ ERA w/o adapt. & 43.69 & 39.15 & 40.16 & 58.66 & 40.65 & 34.49 & 47.84 \\
     & \textbf{Asym.\ ERA} & \textbf{45.70} & \textbf{42.01} & \textbf{44.20} & 59.33 & \textbf{44.13} & \textbf{37.16} & \textbf{48.59} \\
    \midrule
    \multirow{6}{*}{OpenAI-small} & Zero-shot & 36.74 & 35.62 & 41.35 & 50.36 & 31.77 & 30.64 & 36.67 \\
     & Embedding Adapter & 20.63 & 17.98 & 26.01 & 21.56 & 17.35 & 17.88 & 21.12 \\
     & Sym.\ ERA w/o adapt. & 37.19 & 36.32 & 41.01 & 50.13 & 30.79 & 31.09 & 37.71 \\
     & Sym.\ ERA & 43.26 & 41.56 & 46.89 & \textbf{56.05} & 35.91 & 36.84 & 44.42 \\
     & Asym.\ ERA w/o adapt. & 46.63 & 41.69 & 49.48 & 52.68 & 34.72 & 40.40 & 50.86 \\
     & \textbf{Asym.\ ERA} & \textbf{48.18} & \textbf{44.57} & \textbf{50.93} & 53.52 & \textbf{38.34} & \textbf{41.65} & \textbf{51.84} \\
    \midrule
    \multirow{6}{*}{Qwen3-0.6B} & Zero-shot & 50.37 & 48.48 & 58.34 & 60.56 & 48.58 & 44.55 & 49.17 \\
     & Embedding Adapter & 28.71 & 23.47 & 39.04 & 37.71 & 22.41 & 22.62 & 28.92 \\
     & Sym.\ ERA w/o adapt. & 51.12 & 49.28 & 57.28 & 62.17 & \textbf{49.85} & 45.53 & 50.20 \\
     & Sym.\ ERA & 51.62 & 49.49 & 57.85 & 62.74 & 49.52 & \textbf{45.96} & 50.95 \\
     & Asym.\ ERA w/o adapt. & 51.44 & 50.11 & 57.53 & 63.13 & 48.35 & 44.70 & 51.07 \\
     & \textbf{Asym.\ ERA} & \textbf{53.02} & \textbf{50.73} & \textbf{59.32} & \textbf{63.27} & 49.45 & 45.21 & \textbf{53.55} \\
    \midrule
    \multirow{4}{*}{Qwen3-8B} & Zero-shot & 57.91 & 56.94 & 61.85 & 69.38 & 57.89 & \textbf{53.75} & 56.65 \\
     & Embedding Adapter & 35.32 & 33.45 & 43.34 & 43.45 & 29.77 & 24.64 & 36.88 \\
     & Sym.\ ERA w/o adapt. & 58.51 & \textbf{58.85} & 61.53 & 70.03 & \textbf{58.25} & 53.39 & 57.56 \\
     & Sym.\ ERA & \textbf{58.92} & 58.61 & \textbf{63.26} & \textbf{70.75} & 57.72 & 53.33 & \textbf{58.03} \\
    \bottomrule
  \end{tabular}
  }
  \caption{nDCG@10 (\%) on the MAIR benchmark with 10\% of labeled query-document pairs. }
  \label{tab:main_results_10percent}
\end{table*}

\begin{table*}[!t]
  \centering
  \small
    \scalebox{0.85}{
  \begin{tabular}{llrrrrrrr}
    \toprule
    Base model & Variant & \textbf{Avg} & Academic & Code & Finance & Legal & Medical & Web \\
    \midrule
    \multirow{6}{*}{BGE-M3} & Zero-shot & 34.03 & 34.81 & 34.85 & 52.44 & 37.01 & 25.25 & 33.28 \\
     & Embedding Adapter & 33.24 & 28.17 & 37.72 & 48.70 & 33.55 & 24.28 & 34.05 \\
     & Sym.\ ERA w/o adapt. & 34.17 & 35.01 & 33.79 & 53.58 & 38.41 & 25.54 & 33.35 \\
     & Sym.\ ERA & 43.65 & 41.18 & 43.27 & \textbf{61.57} & 43.39 & 34.46 & 45.15 \\
     & Asym.\ ERA w/o adapt. & 43.69 & 39.15 & 40.16 & 58.66 & 40.65 & 34.49 & 47.84 \\
     & \textbf{Asym.\ ERA} & \textbf{47.27} & \textbf{42.91} & \textbf{47.89} & 60.50 & \textbf{48.03} & \textbf{38.05} & \textbf{49.49} \\
    \midrule
    \multirow{6}{*}{OpenAI-small} & Zero-shot & 36.74 & 35.62 & 41.35 & 50.36 & 31.77 & 30.64 & 36.67 \\
     & Embedding Adapter & 37.21 & 30.74 & 42.68 & 47.38 & 31.47 & 32.23 & 38.71 \\
     & Sym.\ ERA w/o adapt. & 37.19 & 36.32 & 41.01 & 50.13 & 30.79 & 31.09 & 37.71 \\
     & Sym.\ ERA & 46.37 & 42.37 & 50.59 & \textbf{57.32} & 39.14 & 39.98 & 48.24 \\
     & Asym.\ ERA w/o adapt. & 46.63 & 41.69 & 49.48 & 52.68 & 34.72 & 40.40 & 50.86 \\
     & \textbf{Asym.\ ERA} & \textbf{50.12} & \textbf{46.25} & \textbf{54.67} & 55.81 & \textbf{42.62} & \textbf{43.40} & \textbf{52.81} \\
    \midrule
    \multirow{6}{*}{Qwen3-0.6B} & Zero-shot & 50.37 & 48.48 & 58.34 & 60.56 & 48.58 & 44.55 & 49.17 \\
     & Embedding Adapter & 43.76 & 38.92 & 54.98 & 54.03 & 39.69 & 36.76 & 43.22 \\
     & Sym.\ ERA w/o adapt. & 51.12 & 49.28 & 57.28 & 62.17 & 49.85 & 45.53 & 50.20 \\
     & Sym.\ ERA & 52.57 & 49.46 & 59.13 & \textbf{64.03} & 51.73 & \textbf{46.85} & 51.80 \\
     & Asym.\ ERA w/o adapt. & 51.44 & 50.11 & 57.53 & 63.13 & 48.35 & 44.70 & 51.07 \\
     & \textbf{Asym.\ ERA} & \textbf{54.03} & \textbf{51.23} & \textbf{59.95} & 63.79 & \textbf{52.60} & 46.64 & \textbf{54.33} \\
    \midrule
    \multirow{4}{*}{Qwen3-8B} & Zero-shot & 57.91 & 56.94 & 61.85 & 69.38 & 57.89 & 53.75 & 56.65 \\
     & Embedding Adapter & 51.79 & 49.51 & 59.85 & 63.16 & 50.43 & 43.10 & 51.42 \\
     & Sym.\ ERA w/o adapt. & 58.51 & 58.85 & 61.53 & 70.03 & 58.25 & 53.39 & 57.56 \\
     & Sym.\ ERA & \textbf{59.99} & \textbf{59.20} & \textbf{64.57} & \textbf{71.82} & \textbf{59.16} & \textbf{54.75} & \textbf{58.96} \\
    \bottomrule
  \end{tabular}
  }
  \caption{nDCG@10 (\%) on the MAIR benchmark with 40\% of labeled query-document pairs. }
  \label{tab:main_results_40percent}
\end{table*}

\subsection{Main Results with Recall@100 and MAP@100}
\label{sssec:main_results_recall_map}
In addition to nDCG@10, we also evaluate the performance of ERA and baselines using Recall@100 and MAP@100, which are shown in Table \ref{tab:main_results_recall} and Table \ref{tab:main_results_map}, respectively.

\begin{table*}[!t]
  \centering
  \small
    \scalebox{0.85}{
  \begin{tabular}{llrrrrrrr}
    \toprule
    Base model & Variant & \textbf{Avg} & Academic & Code & Finance & Legal & Medical & Web \\
    \midrule
    \multirow{6}{*}{BGE-M3} & Zero-shot & 55.50 & 69.34 & 68.05 & 84.97 & 58.20 & 20.99 & 54.44 \\
     & Embedding Adapter & 51.49 & 63.09 & 65.85 & 75.59 & 56.98 & 19.53 & 49.82 \\
     & Sym.\ ERA w/o adapt. & 55.28 & 70.35 & 66.55 & 84.28 & 58.43 & 20.76 & 54.27 \\
     & Sym.\ ERA & 64.17 & 78.53 & 73.59 & 89.03 & 66.86 & 26.56 & 65.78 \\
     & Asym.\ ERA w/o adapt. & 65.91 & 76.80 & 79.19 & 89.52 & 65.44 & 26.14 & 68.86 \\
     & \textbf{Asym.\ ERA} & \textbf{68.62} & \textbf{80.58} & \textbf{81.22} & \textbf{90.20} & \textbf{69.77} & \textbf{30.65} & \textbf{70.81} \\
    \midrule
    \multirow{6}{*}{OpenAI-small} & Zero-shot & 60.99 & 74.67 & 75.96 & 88.69 & 62.11 & 27.82 & 59.29 \\
     & Embedding Adapter & 57.71 & 65.21 & 73.27 & 81.62 & 60.62 & 27.14 & 56.91 \\
     & Sym.\ ERA w/o adapt. & 61.18 & 75.88 & 75.26 & 87.99 & 61.02 & 27.01 & 60.23 \\
     & Sym.\ ERA & 67.83 & 81.18 & 81.18 & 92.02 & 69.25 & 32.90 & 67.85 \\
     & Asym.\ ERA w/o adapt. & 68.62 & 79.34 & 83.84 & 91.87 & 66.55 & 33.35 & 69.75 \\
     & \textbf{Asym.\ ERA} & \textbf{70.98} & \textbf{83.92} & \textbf{84.50} & \textbf{93.52} & \textbf{71.59} & \textbf{36.00} & \textbf{71.48} \\
    \midrule
    \multirow{6}{*}{Qwen3-0.6B} & Zero-shot & 70.24 & 82.39 & 88.53 & 91.87 & 71.11 & 36.58 & 69.01 \\
     & Embedding Adapter & 63.52 & 75.18 & 81.86 & 85.36 & 66.96 & 28.84 & 62.22 \\
     & Sym.\ ERA w/o adapt. & 70.44 & 83.05 & 88.10 & 91.54 & 70.99 & 36.46 & 69.54 \\
     & Sym.\ ERA & 72.00 & 84.48 & 88.03 & 91.77 & 73.48 & \textbf{37.39} & 71.90 \\
     & Asym.\ ERA w/o adapt. & 71.33 & 83.95 & 88.31 & 93.35 & 71.97 & 35.24 & 71.24 \\
     & \textbf{Asym.\ ERA} & \textbf{73.56} & \textbf{86.31} & \textbf{89.02} & \textbf{93.51} & \textbf{74.06} & 37.13 & \textbf{74.39} \\
    \midrule
    \multirow{4}{*}{Qwen3-8B} & Zero-shot & 76.22 & 87.62 & 90.57 & 96.69 & 76.78 & \textbf{43.74} & 76.34 \\
     & Embedding Adapter & 70.24 & 81.84 & 85.47 & 93.12 & 72.58 & 34.82 & 70.31 \\
     & Sym.\ ERA w/o adapt. & 76.36 & 87.57 & 90.42 & 97.46 & 76.83 & 42.77 & 76.95 \\
     & Sym.\ ERA & \textbf{77.08} & \textbf{88.10} & \textbf{90.73} & \textbf{97.87} & \textbf{79.05} & 43.06 & \textbf{77.76} \\
    \bottomrule
  \end{tabular}
  }
  \caption{Recall@100 (\%) on the MAIR benchmark with 20\% of labeled query-document pairs. }
  \label{tab:main_results_recall}
\end{table*}

\begin{table*}[!t]
  \centering
  \small
    \scalebox{0.85}{
  \begin{tabular}{llrrrrrrr}
    \toprule
    Base model & Variant & \textbf{Avg} & Academic & Code & Finance & Legal & Medical & Web \\
    \midrule
    \multirow{6}{*}{BGE-M3} & Zero-shot & 26.93 & 30.23 & 29.99 & 47.80 & 30.66 & 11.83 & 26.39 \\
     & Embedding Adapter & 19.31 & 17.34 & 25.66 & 37.17 & 21.29 & 7.43 & 18.91 \\
     & Sym.\ ERA w/o adapt. & 27.01 & 30.49 & 29.40 & 49.03 & 30.22 & 11.70 & 26.66 \\
     & Sym.\ ERA & 33.29 & 35.28 & 35.05 & \textbf{56.08} & \textbf{34.70} & 14.60 & 35.03 \\
     & Asym.\ ERA w/o adapt. & 33.34 & 33.43 & 34.43 & 53.55 & 30.05 & 13.44 & 37.63 \\
     & \textbf{Asym.\ ERA} & \textbf{36.20} & \textbf{36.70} & \textbf{40.34} & 54.55 & 33.94 & \textbf{16.20} & \textbf{39.44} \\
    \midrule
    \multirow{6}{*}{OpenAI-small} & Zero-shot & 28.97 & 30.25 & 36.35 & 46.19 & 27.65 & 14.12 & 29.07 \\
     & Embedding Adapter & 22.04 & 19.19 & 29.29 & 33.80 & 20.66 & 11.96 & 22.56 \\
     & Sym.\ ERA w/o adapt. & 29.29 & 30.83 & 36.05 & 46.01 & 26.59 & 13.43 & 30.23 \\
     & Sym.\ ERA & 35.80 & 36.43 & 42.58 & \textbf{52.50} & 32.07 & 17.81 & 37.97 \\
     & Asym.\ ERA w/o adapt. & 36.70 & 35.48 & 43.96 & 48.46 & 30.15 & 18.25 & 40.74 \\
     & \textbf{Asym.\ ERA} & \textbf{39.22} & \textbf{39.15} & \textbf{47.22} & 50.17 & \textbf{35.11} & \textbf{19.76} & \textbf{42.64} \\
    \midrule
    \multirow{6}{*}{Qwen3-0.6B} & Zero-shot & 39.97 & 42.48 & 52.62 & 55.83 & 39.11 & 21.88 & 39.21 \\
     & Embedding Adapter & 29.90 & 28.59 & 43.32 & 46.57 & 27.99 & 13.51 & 29.50 \\
     & Sym.\ ERA w/o adapt. & 40.59 & 43.26 & 52.00 & 57.76 & 39.40 & 21.61 & 40.38 \\
     & Sym.\ ERA & 41.88 & 43.60 & 53.62 & 58.77 & \textbf{40.68} & \textbf{22.07} & 42.17 \\
     & Asym.\ ERA w/o adapt. & 40.71 & 44.01 & 51.78 & 58.45 & 37.87 & 20.61 & 41.06 \\
     & \textbf{Asym.\ ERA} & \textbf{42.70} & \textbf{44.58} & \textbf{54.14} & \textbf{59.11} & 39.89 & 21.51 & \textbf{43.92} \\
    \midrule
    \multirow{4}{*}{Qwen3-8B} & Zero-shot & 47.08 & 50.76 & 56.21 & 63.82 & 47.24 & \textbf{28.38} & 47.02 \\
     & Embedding Adapter & 36.45 & 36.72 & 47.69 & 52.14 & 35.39 & 16.64 & 37.50 \\
     & Sym.\ ERA w/o adapt. & 47.65 & 52.53 & 56.01 & 64.75 & 47.02 & 27.21 & 48.21 \\
     & Sym.\ ERA & \textbf{48.55} & \textbf{52.56} & \textbf{57.80} & \textbf{66.99} & \textbf{47.64} & 27.14 & \textbf{49.26} \\
    \bottomrule
  \end{tabular}
  }
  \caption{MAP@100 (\%) on the MAIR benchmark with 20\% of labeled query-document pairs. }
  \label{tab:main_results_map}
\end{table*}

\subsection{Domain Generality with Varying Train Ratios}
\label{app:ssec:domain_generality_train_ratios}

\subsubsection{Out-of-Domain Experiment.}
\begin{figure*}[!t]
   \centering
   \begin{subfigure}[b]{0.32\linewidth}
      \centering
      \includegraphics[width=\linewidth]{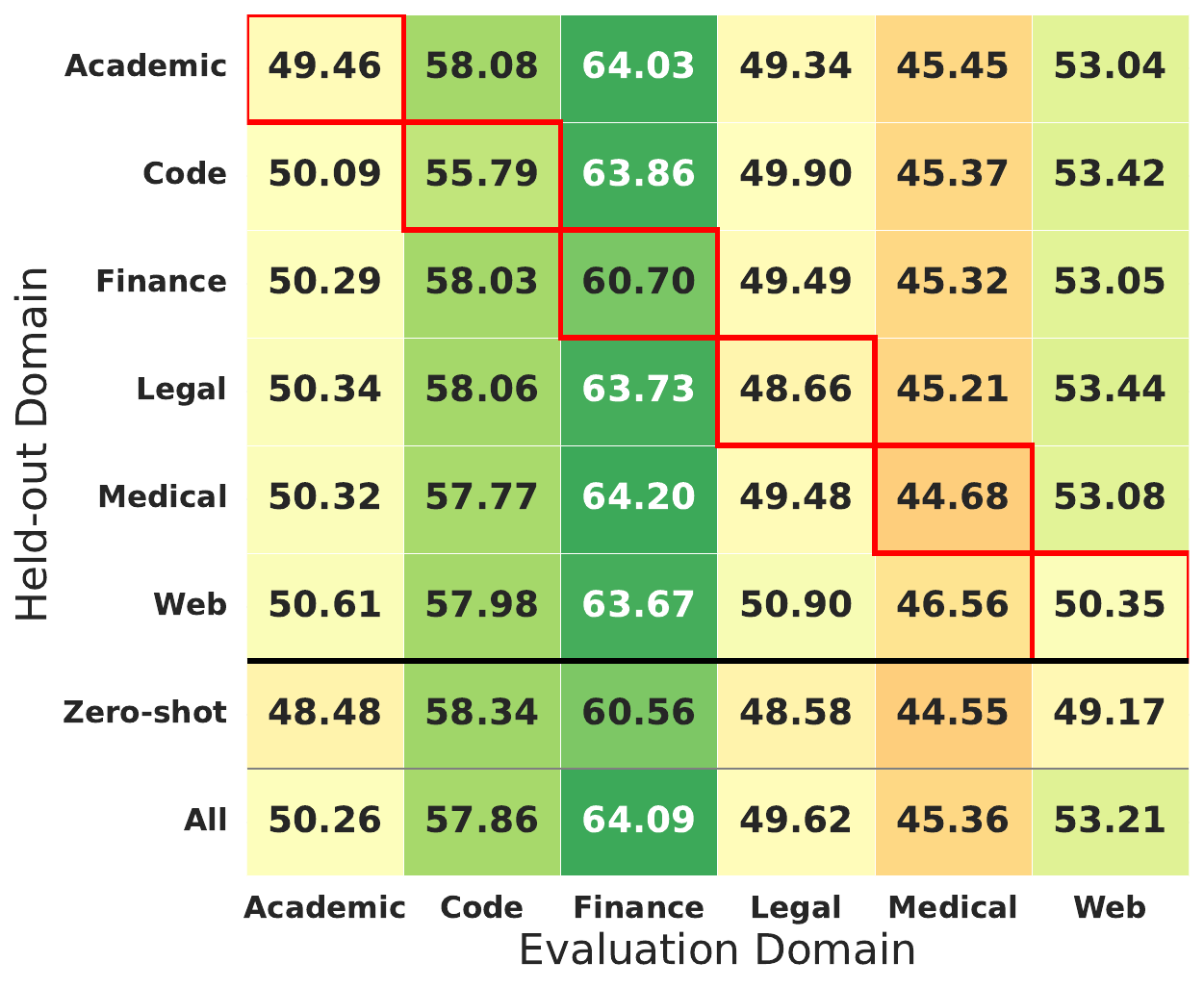}
      \caption{5\% train ratio}
      \label{fg:app:out_of_domain_5percent}
   \end{subfigure}
   \hfill
   \begin{subfigure}[b]{0.32\linewidth}
      \centering
      \includegraphics[width=\linewidth]{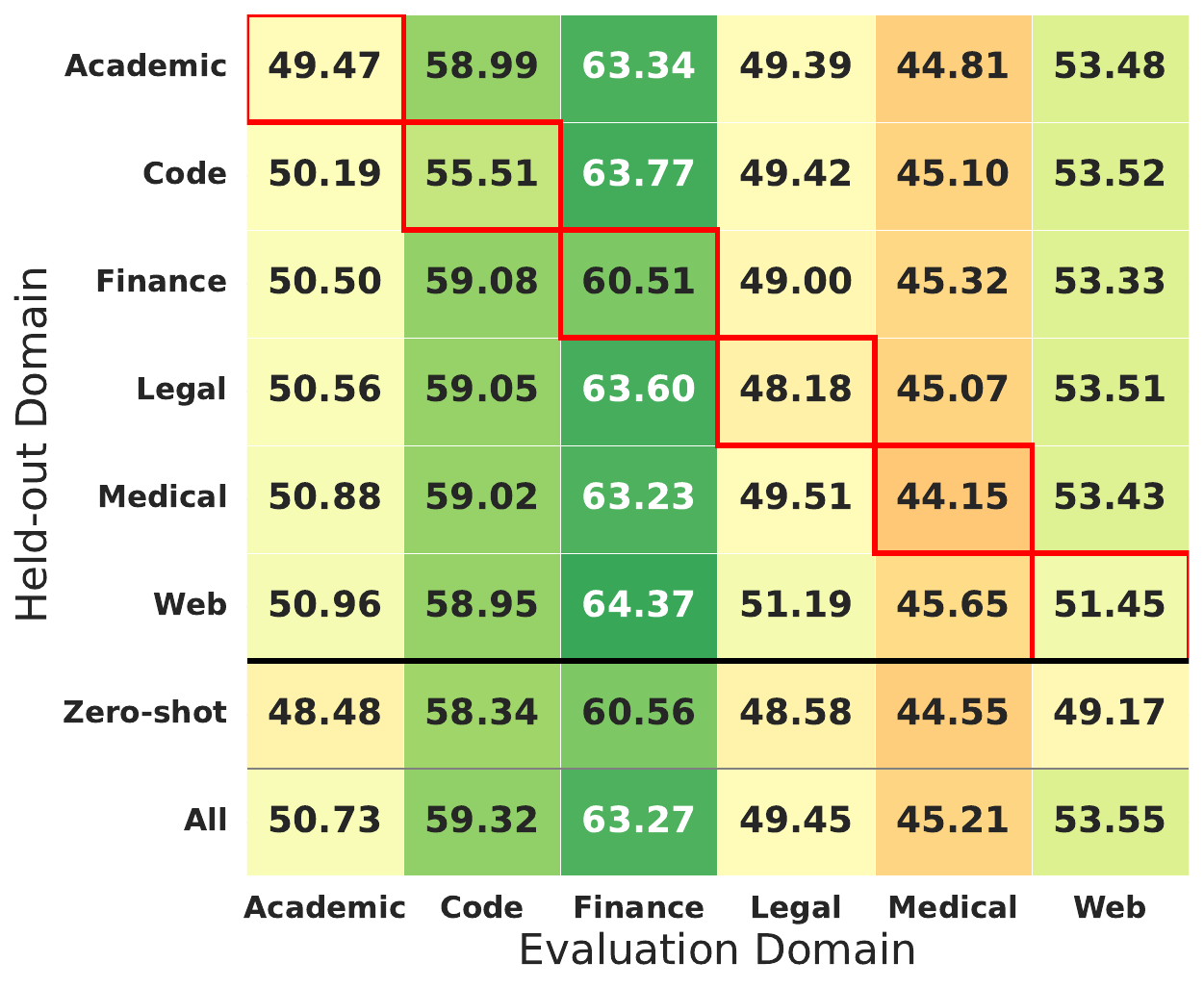}
      \caption{10\% train ratio}
      \label{fg:app:out_of_domain_10percent}
   \end{subfigure}
   \hfill
   \begin{subfigure}[b]{0.32\linewidth}
      \centering
      \includegraphics[width=\linewidth]{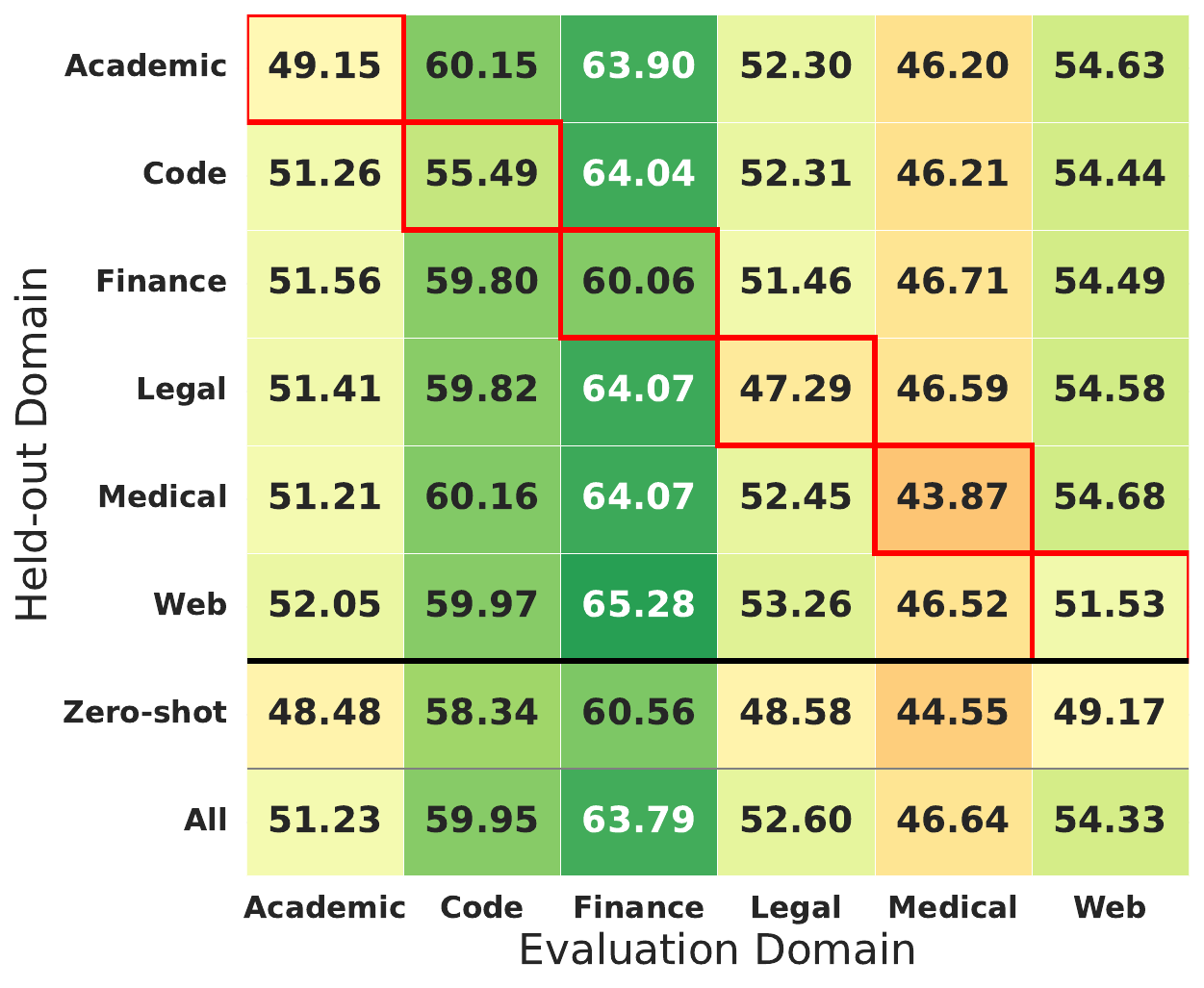}
      \caption{40\% train ratio}
      \label{fg:app:out_of_domain_40percent}
   \end{subfigure}
   \caption{nDCG@10 comparison between zero-shot, ERA trained on all domains, and ERA trained on all domains except the target domain with varying train ratios, where we use Qwen3-8B as the query embedder and Qwen3-0.6B as the document embedder.}
   \label{fg:app:out_of_domain}
\end{figure*}

We show the nDCG@10 results of domain generality analysis with varying percentages of labeled query-document pairs for training, including 5\%, 10\%, and 40\% in Figures \ref{fg:app:out_of_domain} and \ref{fg:app:domain_specific}, respectively.
We observe the same trend that ERA trained on all domains except the target domain shows comparable or slightly lower performance than zero-shot retrieval for the target domain while largely improving the performance in other domains.

\subsubsection{Domain-specific Experiment.}
\begin{figure*}[!t]
   \centering
   \begin{subfigure}[b]{0.32\linewidth}
      \centering
      \includegraphics[width=\linewidth]{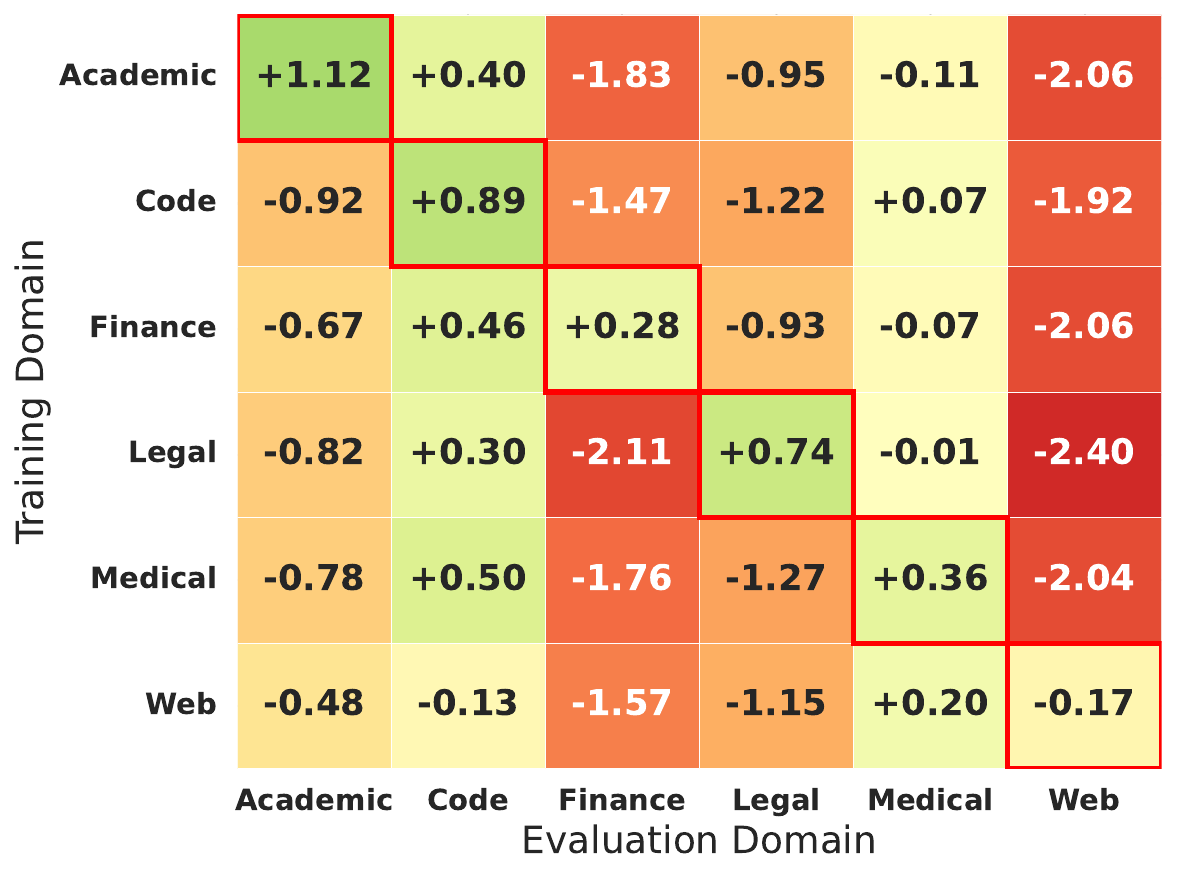}
      \caption{5\% train ratio}
      \label{fg:app:domain_specific_5percent}
   \end{subfigure}
   \hfill
   \begin{subfigure}[b]{0.32\linewidth}
      \centering
      \includegraphics[width=\linewidth]{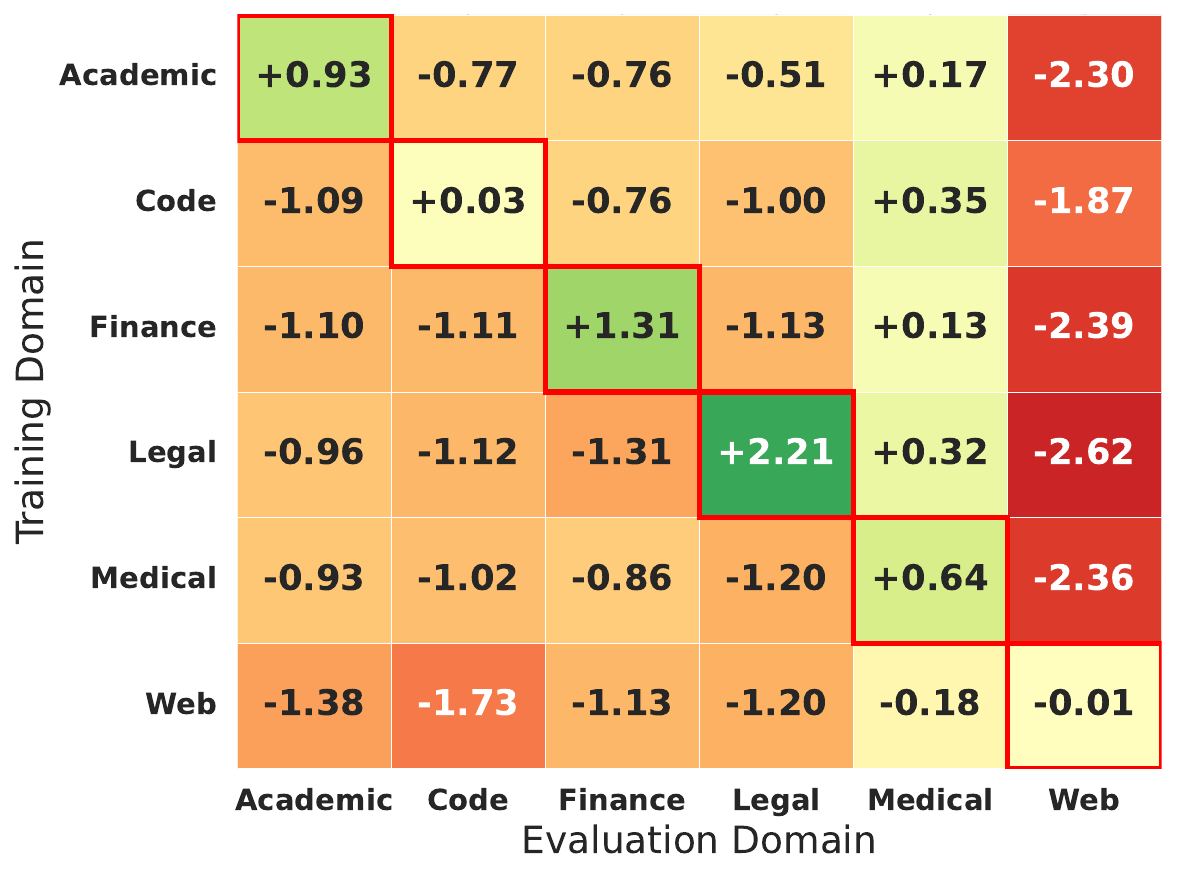}
      \caption{10\% train ratio}
      \label{fg:app:domain_specific_10percent}
   \end{subfigure}
   \hfill
   \begin{subfigure}[b]{0.32\linewidth}
      \centering
      \includegraphics[width=\linewidth]{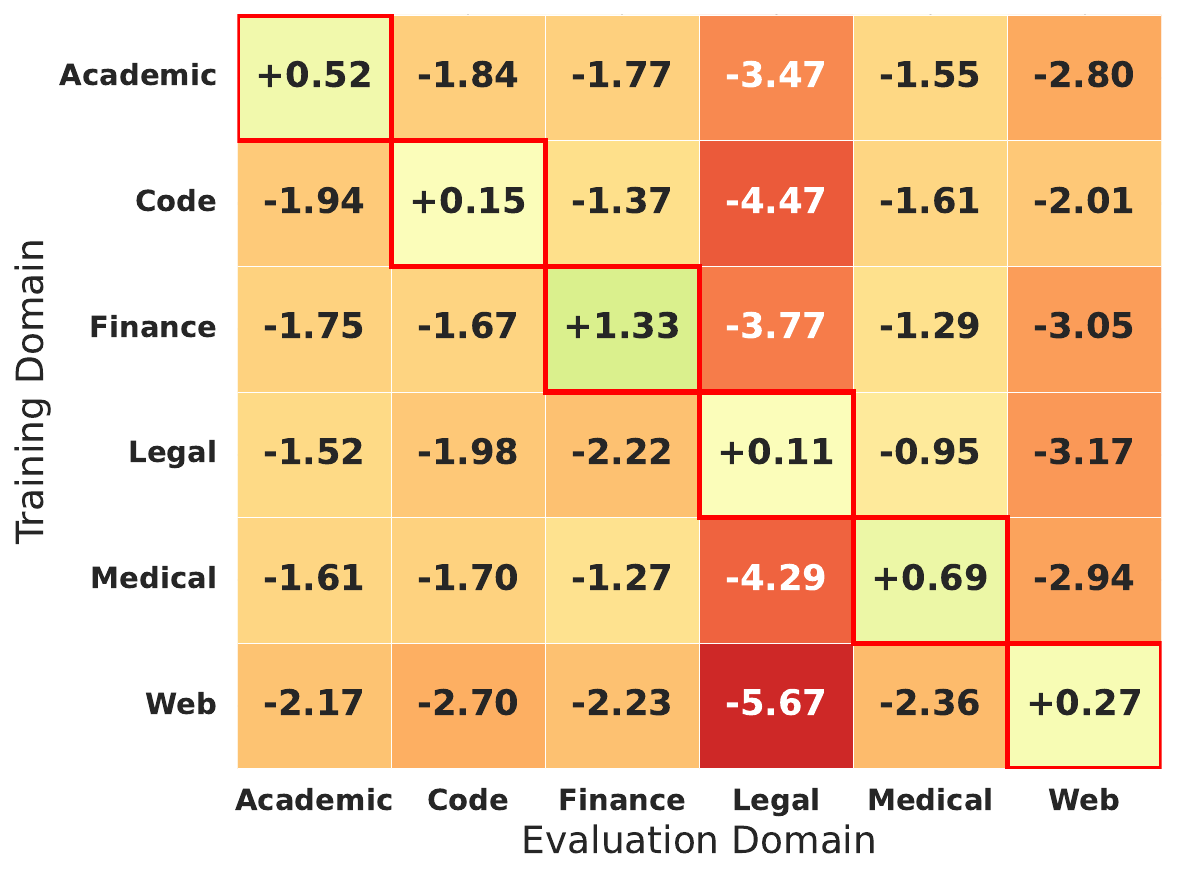}
      \caption{40\% train ratio}
      \label{fg:app:domain_specific_40percent}
   \end{subfigure}
   \caption{nDCG@10 difference between ERA trained on all domains and ERA trained on the target domain with varying train ratios, where we use Qwen3-8B as the query embedder and Qwen3-0.6B as the document embedder.}
   \label{fg:app:domain_specific}
\end{figure*}

We show the nDCG@10 difference between ERA trained on all domains and ERA trained on the target domain with varying percentages of labeled query-document pairs for training, including 5\%, 10\%, and 40\% in Figures \ref{fg:app:domain_specific_5percent}, \ref{fg:app:domain_specific_10percent}, and \ref{fg:app:domain_specific_40percent}, respectively.
The experimental results show the same trend that while the performance of domain-specific adapters trained on only the target domain can be slightly better than the general adapter trained on each domain where they are trained, the performance difference is generally small across all domains.

\subsection{Hyperparameter Search Results}
\label{sssec:hp_search_results}
We show the hyperparameter search results for the learning rate and weight decay of the adaptation stage in Table \ref{tab:hp_search}.
We observe that the performance is generally stable, i.e., 0.27\% nDCG@10 difference between the best and worst configurations, which suggests that ERA is not sensitive to the choice of these hyperparameters.

\begin{table*}[t]
  \centering
  \small
  \begin{tabular}{lrrrrr}
    \toprule
    WD $\setminus$ LR & $10^{-5}$ & $5\times10^{-5}$ & $10^{-4}$ & $5\times10^{-4}$ & $10^{-3}$ \\
    \midrule
    $10^{-2}$ & 52.83 & 52.59 & 52.60 & 52.58 & 52.60 \\
    $10^{-3}$ & 52.83 & 52.59 & 52.57 & 52.57 & 52.58 \\
    $10^{-4}$ & \textbf{52.84} & 52.59 & 52.56 & 52.57 & 52.60 \\
    \bottomrule
  \end{tabular}
  \caption{Hyperparameter search results (nDCG@10 \%). Rows: weight decay (WD), columns: learning rate (LR). \textbf{Bold} = best configuration.}
  \label{tab:hp_search}  
\end{table*}

\cameraready{
\subsection{Results on Non-instructed Retrieval Benchmark}
\label{app:ssec:beir}
To examine whether ERA's effectiveness depends on the instruction-rich setting, e.g., MAIR,
we additionally evaluate ERA on MS MARCO from the BEIR benchmark \cite{thakur2021beir}, whose queries contain no task instructions. 
We use Qwen3-0.6B for a document embedder and Qwen3-8B for a query embedder in the asymmetric setting and employed the linear weight trained from MAIR with a 40\% train ratio, i.e., we do not train the weight on MS MARCO.

\begin{table}[t]
  \centering
  \resizebox{\linewidth}{!}{%
  \begin{tabular}{lrrr}
    \toprule
    Method & nDCG@10 & Recall@100 & MAP@100 \\
    \midrule
    Zero-shot & 26.30 & 72.83 & 22.29 \\
    \textbf{ERA} (0.6B $\rightarrow$ 0.6B) & 32.95 & \textbf{82.97} & 28.16 \\
    \textbf{ERA} (8B $\rightarrow$ 0.6B) & \textbf{33.52} & 82.91 & \textbf{28.74} \\
    \bottomrule
  \end{tabular}%
  }
  \caption{Retrieval performance (\%) on MS MARCO \texttt{dev} (8.8M passages, exact cosine top-100 search). ERA is trained on MAIR and applied to MS MARCO without any in-domain training.}
  \label{tab:msmarco_main_results}
\end{table}

Table \ref{tab:msmarco_main_results} shows that ERA achieves better performance than zero-shot in the MS MARCO dataset. This is a similar trend to the results on the MAIR benchmark. 
The result indicates ERA is helpful for not only instructed retrieval but non-instructed retrieval.

\subsection{Comparison with Exact Alignment}
\label{app:ssec:exact_alignment}
ERA's alignment stage is intended to provide a retrieval-compatible initialization, rather than to reconstruct the document embedding space pointwise. We therefore compare ERA's learned alignment with an exact alignment solution that directly minimizes the discrepancy between paired embeddings of identical texts.
To this end, we additionally test closed-form Orthogonal Procrustes alignment. 
Given paired embeddings $X$ and $Y$ of the same texts, let $\mu_X$ and $\mu_Y$ denote their respective mean vectors, and define the centered embeddings as $X_c = X - \mu_X$ and $Y_c = Y - \mu_Y$. We then compute

$$
X_c^\top Y_c = U\Sigma V^\top,\qquad R = UV^\top.
$$
The aligned embedding is
$$
\hat{y}=(x-\mu_X)R+\mu_Y.
$$

Because the symmetric setting uses the same 1024-dimensional Qwen3-0.6B model on both sides, $R$ is a square orthogonal matrix. We store this transformation as a standard linear adapter, with weight $R^\top$ and bias $\mu_Y-\mu_XR$, and apply the same output normalization as ERA. For the label-training experiment, we simply use the resulting Procrustes adapter as the initialization and then run the same label-training pipeline as ERA. We conduct a hyperparameter search for the label-training with the same space of weight decay and learning rate for the main experiments and used the best parameter. 

\begin{table}[t]
  \centering
  \resizebox{\linewidth}{!}{
  \begin{tabular}{lrrr}
    \toprule
    Method & nDCG@10 & R@100 & MAP@100 \\
    \midrule
    Procrustes w/o adapt. & 51.13 & 70.43 & 40.59 \\
    Sym. ERA w/o adapt. & 51.12 & 70.44 & 40.59 \\
    Procrustes label training & 51.58 & 71.72 & 41.25 \\
    \textbf{Sym. ERA} & \textbf{51.62} & \textbf{71.75} & \textbf{41.29} \\
    \bottomrule
  \end{tabular}
  }
  \caption{Comparison of the closed-form alignment baseline. The train ratio for labal training is set to 10\%. }
  \label{tab:closed_form_alignment}
\end{table}

Table \ref{tab:closed_form_alignment} shows the results comparing closed-form alignment against ERA Our investigation reveals that Procrustes alignment without refinement yields retrieval performance nearly identical to that of ERA's initial alignment phase. Nevertheless, the full ERA framework demonstrates a marginal performance advantage following the supervised adaptation stage. We hypothesize that employing a consistent optimization objective across both the alignment and adaptation phases facilitates more robust parameter refinement, which is particularly advantageous in low-label settings.

Classical Orthogonal Procrustes requires the two embedding spaces to have the same dimensionality. For different-dimensional embeddings, alternatives such as affine ridge regression could be considered; we leave their investigation to future work.

\subsection{Combination with Reranking}
\label{app:ssec:reranking}
ERA operates at the first-stage retrieval level and is therefore orthogonal to downstream reranking. To test whether its gains persist in a standard two-stage retrieval pipeline, we apply the same reranker to the candidates produced by the zero-shot retriever and ERA.
Specifically, we use a strong reranker, Qwen3-reranker-8B \cite{zhang2025qwen3embedding}, after retrieving top-100 documents by baselines and ERA. Other settings are the same as Table \ref{tab:main_results} in the paper.

Table \ref{tab:reranker_main} shows the results. 
We observe that reranking improves all methods and Asymmetric ERA with reranking achieves the best nDCG@10 in both BGE-M3 and Qwen3-0.6B. This indicates that ERA and reranking are orthogonal and combining them provides further improvement over using only one of them.

\begin{table}[t]
  \centering
  \resizebox{\linewidth}{!}{%
  \begin{tabular}{llrr}
    \toprule
    Base model & Variant & Before & Reranked \\
    \midrule
    \multirow{6}{*}{BGE-M3} & Zero-shot & 34.14 & 53.49 \\
     & Embedding Adapter & 26.18 & 50.04 \\
     & Sym.\ ERA w/o adapt. & 34.17 & 53.49 \\
     & Sym.\ ERA & 42.14 & 59.29 \\
     & Asym.\ ERA w/o adapt. & 43.68 & 60.12 \\
     & \textbf{Asym.\ ERA} & \textbf{46.59} & \textbf{61.86} \\
    \midrule
    \multirow{6}{*}{Qwen3-0.6B} & Zero-shot & 51.12 & 62.01 \\
     & Embedding Adapter & 38.94 & 58.70 \\
     & Sym.\ ERA w/o adapt. & 51.11 & 62.02 \\
     & Sym.\ ERA & 52.19 & 62.60 \\
     & Asym.\ ERA w/o adapt. & 51.44 & 62.93 \\
     & \textbf{Asym.\ ERA} & \textbf{53.34} & \textbf{63.68} \\
    \bottomrule
  \end{tabular}%
  }
  \caption{nDCG@10 (\%) on the MAIR benchmark before and after reranking with Qwen3-Reranker-8B. Top-100 first-stage candidates are reranked. \textbf{Bold} indicates the best result for each base model and column. Asymmetric ERA uses Qwen3-8B as the query encoder and the base model as the document encoder.}
  \label{tab:reranker_main}
\end{table}

}

\end{document}